\documentclass[10pt,letterpaper]{article}
\usepackage[top=0.85in,left=2.75in,footskip=0.75in]{geometry}

\usepackage{ulem}
\newcommand\msout{\bgroup\markoverwith{\textcolor{blue}{\rule[0.5ex]{1pt}{1pt}}}\ULon} 
\newcommand\msoutv{\bgroup\markoverwith{\textcolor{red}{\rule[0.5ex]{1pt}{1pt}}}\ULon} 

\usepackage{amsmath,amssymb}

\usepackage{changepage}
\usepackage{float}
\usepackage[utf8x]{inputenc}

\usepackage{textcomp,marvosym}

\usepackage{cite}

\usepackage{nameref,hyperref}

\usepackage{microtype}
\DisableLigatures[f]{encoding = *, family = * }

\usepackage[final]{pdfpages}

\usepackage{array}

\newcolumntype{+}{!{\vrule width 2pt}}

\newlength\savedwidth

\raggedright
\usepackage[aboveskip=1pt,labelfont=bf,labelsep=period,justification=raggedright,singlelinecheck=off]{caption}

\makeatletter
\renewcommand{\@biblabel}[1]{\quad#1.}
\makeatother

\usepackage{lastpage,fancyhdr,graphicx}
\usepackage{epstopdf}
\fancyheadoffset[L]{2.25in}
\fancyfootoffset[L]{2.25in}
\begin{document}
\vspace*{0.2in}

\begin{flushleft}
{\Large
\textbf\newline{Modeling viral mutations in the spread of epidemics} 
}
\newline
\\
Vitor M. Marquioni\textsuperscript{1},
Marcus A. M. de Aguiar\textsuperscript{1*}
\\
\bigskip
\textbf{$^1$} Instituto de Física ``Gleb Wataghin'', Universidade Estadual de Campinas -  UNICAMP, Campinas, SP, Brazil
\\
\bigskip

%
%



* aguiar@ifi.unicamp.br

\end{flushleft}
\section*{Abstract}
Although traditional models of epidemic spreading focus on the number of infected, susceptible and recovered individuals, a lot of attention has been devoted to integrate epidemic models with population genetics. Here we develop an individual-based model for epidemic spreading on networks in which viruses are explicitly represented by finite chains of nucleotides that can mutate inside the host. Under the hypothesis of neutral evolution  we compute analytically the average pairwise genetic distance between all infecting viruses over time. We also derive a mean-field version of this equation that can be added directly to compartmental models such as SIR or SEIR to estimate the genetic evolution. We compare our results with the inferred genetic evolution of SARS-CoV-2 at the beginning of the epidemic in China and found good agreement with the analytical solution of our model. Finally, using genetic distance as a proxy for different strains, we use numerical simulations to show that the lower the connectivity between communities, e.g., cities, the higher the probability of reinfection.

\section*{Author summary}

In this work we describe the genetic evolution of viruses in the course of an epidemic. The viruses are described by their RNA, modeled as a finite sequence of loci with four possible entries representing nucleotides. Viruses mutate at a fixed rate and we assume that genetic variations do not confer differential fitness,  meaning that infected individuals acquire perfect cross immunity against all viral strains. Individuals in the population are represented by nodes of a network of contacts. We compute the diversity of viral population, measured by genetic distance between viral sequences, defined as number of loci bearing different nucleotides. We derive an equation for the evolution of the average genetic distance that depends only on epidemic variables, such as the number of infected and recovered individuals, number of nucleotides and mutation rate. We apply this equation to the beginning of SARS-CoV-2 epidemic in China and show that it agrees well with the available data. We also show how the genetic variability is affected when the virus spreads over connected communities, influencing the probabilities of reinfection.


\section*{\label{sec:introduction}Introduction}

In the late 2019, the world saw the emergence of a new disease, caused by a new type of coronavirus\cite{zhu2020novel} which can cause severe injures to human respiratory system.\cite{wiersinga2020pathophysiology} Since then, we witnessed an uninterrupted worldwide effort in the search for efficient treatments\cite{wiersinga2020pathophysiology,sanders2020pharmacologic}, vaccines\cite{lurie2020developing,le2020covid,graham2020rapid} and better understanding of the epidemic parameters and its pathways of spread\cite{backer2020incubation,wu2020estimating,linton2020incubation,petropoulos2020forecasting}. 

A great number of SARS-CoV-2 genomes has been sequenced in different countries and regions, allowing scientists to study its genealogy and geographic origins\cite{of2020species}.  Different strains have been characterized\cite{forster2020phylogenetic,van2020emergence}, revealing cases of reinfection\cite{to2020covid,tillett2020genomic}. Understanding the mechanisms of mutation and variability in viruses is of utmost importance to forecast forthcoming challenges, e.g. the appearance of other infectious strains or loss of acquired immunity. Mutation rates are usually high in RNA viruses\cite{duffy2018rna} and are important mechanisms for spillover events \cite{duffy2018rna,froissart2010virulence,hawley2013parallel}. Although mutations can have significant impact on the virus genetic machinery, leading to more or less infectious strains\cite{stacey2013eco,de2003mean},  neutral mutations also occur in non-coding RNA regions or if they result in synonymous changes, that do not alter the corresponding protein. Counting the number of mutations and tracking their spread in the population is important for tracing pandemic routes through  communities (neighborhoods, cities, or countries) and giving clues as to how the virus is moving\cite{kupferschmidt2020genome}.

{ Mathematical models of epidemic spreading are crucial to project how the disease will progress and plan intervention strategies, especially in the case of COVID-19 \cite{cobey2004,scabini2020,vasconcelos2020,flaxman2020}.} The great majority of epidemic models divide the population into categories, such as susceptible and infected individuals \cite{kermack1927contribution,anderson1988epidemiology}. Details concerning population structure and how different individuals respond to the infection are ignored, allowing the epidemic spreading to be described by differential equations that can be readily interpreted and solved numerically \cite{keeling2005networks}. The SIR model, susceptible-infectious-recovered, is a classic example of this type of simplification and has set the foundations for the development of more detailed descriptions \cite{kermack1927contribution}. Important extensions include time dependent contact rates \cite{kuznetsov1994bifurcation} and multiple infectious stages occurring in parallel \cite{korobeinikov2009global}. 

One important drawback of the SIR and other related compartmental models is their inability to describe heterogeneity in individual behavior and response to the infection. Some of these features can be introduced with the help of network theory, which provides a framework for modeling explicit population structures \cite{keeling2005networks}. A number of important results were demonstrated in this context, particularly in connection with the distribution of number of contacts among individuals \cite{pastor2001epidemic}. The representation of individuals as nodes of a network can also be combined with stochastic infection and recovering processes, which might have important consequences for viral diversity \cite{buckee2004effects}.

More recently, efforts have been devoted to integrate models of epidemic spreading with population genetics through coalescent theory \cite{slatkin1991pairwise}. This allowed the study of pairwise genetic differences between viral haplotypes, estimation of the viral growth rate \cite{slatkin1991pairwise,volz2009phylodynamics} and times to most recent common ancestor \cite{griffiths1994ancestral,de2016scotti}. Genetic diversity has also been estimated by replacing birth-death models by deterministic epidemic equations \cite{volz2012complex} or introducing population structure \cite{gordo2009genetic}. Multi-strain models were also used to describe how epidemics shape pathogen diversity \cite{kucharski2016capturing}, considering different sources of heterogeneity, such as genotype networks \cite{williams2021localization} or, as we do here, the structure of the host' contact network \cite{buckee2004effects,buckee2007host}.



Here we consider an individual-based model for epidemic spreading where the population is represented by nodes of a network and viruses are modeled explicitly by a binary chain representing their RNA.  This allows us to combine population structure using network theory, stochastic dynamics of epidemic spreading and population genetics into a single framework. One of the advantages of this formulation is that important epidemic features, such as the structure of social contacts through which contamination occurs, viral transmission rates, individual incubation and recover periods, virus's genome length and mutation rate can be readily included and analysed. 

Although many studies have considered imperfect cross-immunity \cite{kucharski2016capturing,williams2021localization,buckee2004effects,buckee2007host}, in the present model we consider only neutral mutations, which do not alter the immune escape or other viral parameters. This implies that, once the host has developed an immune response against a viral strain, it will have perfect cross immunity against all strains. We also assume that all viruses replicating inside the same host are identical, thus they can be modeled by a single RNA sequence. Viruses of two different hosts, however, can be different due to the mutations that happen randomly and independently at each nucleotide. These assumptions are justified if the periods of incubation and sickness are much shorter than the inverse of the mutation rate and the duration of the epidemic.

We track the spreading of the virus through the population network and compute its diversity by tracking the genetic distance between pairs of viruses along the epidemic propagation.  Within this framework, it is possible to study the viral dynamics along different population structures, by changing only the contact network, which is suitable for computational experiments. As an application, we show that the connectivity among different communities (represented by modules of a larger network) changes significantly the viral pairwise distance distribution, suggesting how reinfections could arise if cross-immunity is lost.

Importantly, we derive a recurrence equation for computing the  average  genetic  distance  among  viruses in the population in terms of the number of susceptible and infected individuals, length of the genome and mutation rate. We also derive a mean-field approximation for this equation that can be added to the usual SIR or SEIR models \cite{murray2007mathematical} to estimate the viral genetic evolution in homogeneous populations. Finally, we compare the genetic distance among viruses obtained theoretically from the recurrence equation to the SARS-CoV-2 genomic data, obtained from Chinese epidemic data during the period from 12/23/2019 to 03/24/2020.

The present work is a follow-up of a recently proposed SEIR model designed to study the effects of quarantine regimes\cite{MARQUIONI2020109999}, from which many parameters are obtained.
The paper is organized as follows: in section {\it\nameref{sec:model}}, we describe the SEIR model on networks and how the virus dynamics work.
In {\it\nameref{sec:results1}} we show how to analytically solve this dynamics for the average genetic distance among viruses. Our solution leads to a discrete equation, which we apply to the  SARS-CoV-2 Chinese epidemic data. Taking the continuous time limit we argue that it can be included as a fourth equation to the classic SIR model, enabling one to infer genetic neutral evolution along an epidemic. The mathematical technique we have used can also be implemented in the case of more compartmentalized models.
In {\it\nameref{sec:results2}}, we simulate epidemic spreading along a chain of linearly connected communities and discuss how the risk of reinfection can be increased when the connectivity among them is decreased. This indicates that pandemics are more likely to yield early reinfections than epidemics. We discuss our conclusions in the Section {\it\nameref{sec:discussion}}.

\section*{The Model}
\label{sec:model}

We consider a SEIR individual based model where individuals are divided into four different compartments: {\it Susceptible}, individuals that can be infected; {\it Exposed}, individuals that are infected but not infectious; {\it Infected}, which can spread the virus by infecting others; and {\it Recovered}, who are recovered from the disease and can no longer be infected. We model the population as a network where nodes represent individuals and links indicate connections between them (linked nodes are also termed first neighbors). Time is discrete and at each step all infected individuals may transmit the disease to their susceptible first neighbors with probability $p_I$. { The infection probability can be calculated as $p_I=R_0/(\tau_{0}D)$, where $\tau_{0}$ is the average time duration of symptoms, $R_0$, is the basic reproduction number and $D$ the average network degree. }
Each exposed individual remains in this condition for a time $\tau$ distributed according to $\mathcal{P}(\tau)$ (see Appendix), after which it becomes infected. Every infected can recover with a probability $p_R=1/\tau_0$ per time step\cite{MARQUIONI2020109999}. 

\begin{figure}[H]
\includegraphics[width=14cm]{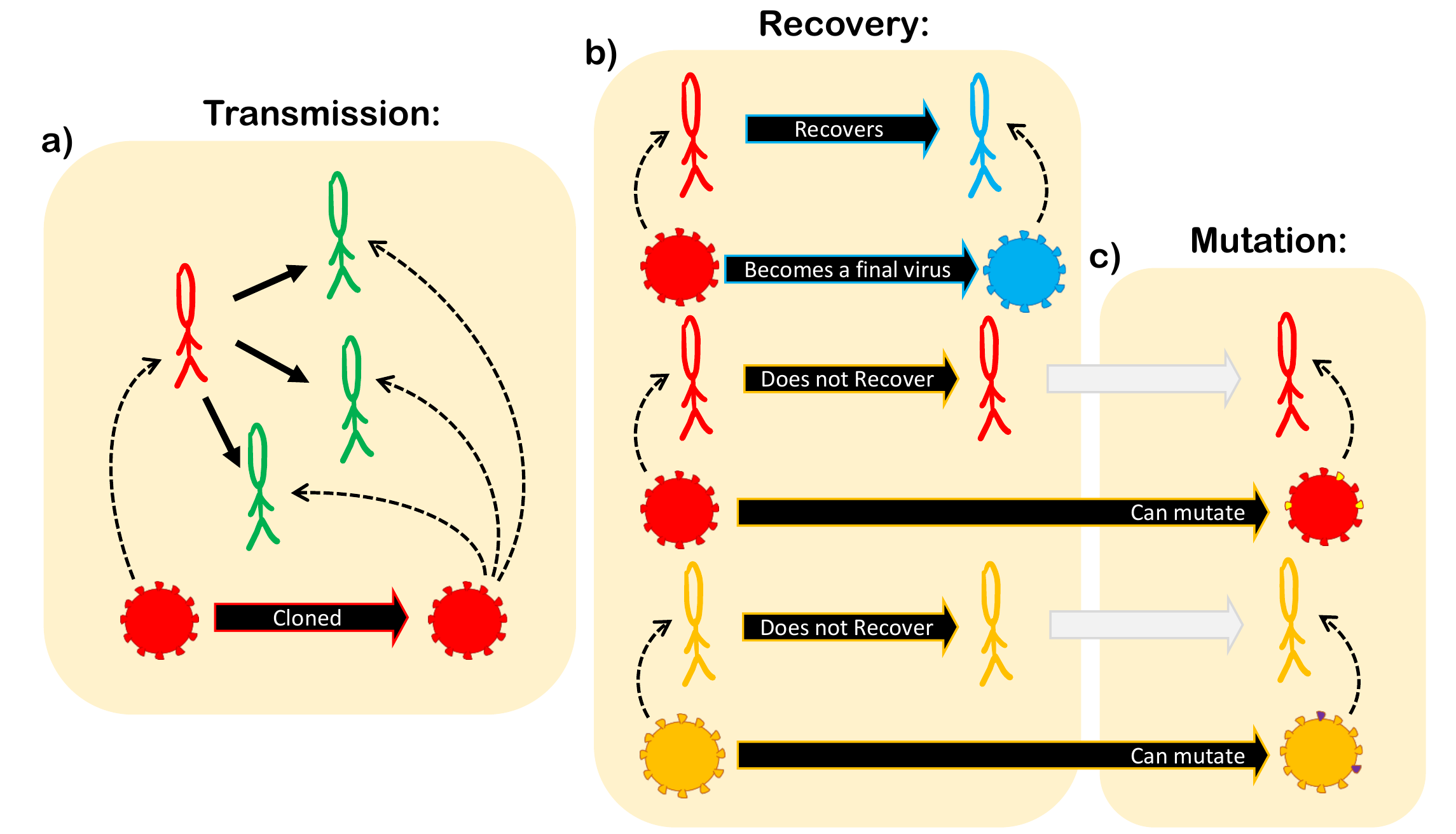}
\caption{{\bf Model dynamics.} {\it (a)} infected individuals (red) can transmit the virus to their susceptible first neighbors (green). When transmission is successful the virus is cloned to the new host, which is now an exposed individual (yellow) and will be able to mutate only in the next iteration. {\it (b)} infected individuals can recover with probability $p_R$. When an individual recovers (blue), its virus stops mutating and becomes a ``final virus.'' {\it (c)} viruses on infected (red) or exposed (yellow) individuals can mutate.}
\label{fig:dynamics}
\end{figure}

Infected and exposed individuals carry a strain of the virus, represented  by a binary chain of size $2B$, where $B$ is the number of nucleotides. Each pair of bits, $b_{2i-1}$ and $b_{2i}$ in the chain ($i=1, \dots, B$) represents a nucleotide, given, for instance, by 00=A, 01=U, 10=C and 11=G. As long as the virus remains hosted in the individual, it can mutate with probability of substitution $\mu$ per nucleotide at every iteration. When the virus is passed from one host to another, it is entirely copied to the new host. When the individual recovers, its virus' RNA stops mutating and its final configuration is saved for further analysis. We call this ``a final virus". Fig.\ref{fig:dynamics} illustrates this dynamics.

To compare the different viruses that appear during the simulation we use the Hamming Distance $d^{\alpha\beta}$, which counts the number of different nucleotides between two viruses $\alpha$ and $\beta$\cite{de2017speciation,sung2009algorithms}. In our model the Hamming distance is given by
\begin{equation}
    d^{\alpha\beta}=B-\sum_{i=1}^{B}\left(|b_{2i-1}^\alpha-b_{2i-1}^\beta|-1\right)\left(|b_{2i}^\alpha-b_{2i}^\beta|-1\right)
\end{equation}
where $b_j^\gamma \in \{0,1\}$  is bit $j$ of the virus $\gamma$. 

We consider a neutral model for the virus evolution and do not include mechanisms of selection. The mutation probability is the same for all nucleotides, independent of its location in the genome or the nitrogenous base the nucleotide changes from or to. Additionally, once an individual { recovers from infection by} a strain it acquires perfect cross immunity against all strains. 


We start the simulation with a single infected individual with genome $b_j^\gamma = 1$ for all $j$. All simulation parameters, can be found in \nameref{appendixes}, { and are scaled so that the time unit is one day}.

\subsection*{\label{sec:results1}Analytical Description}

The analysis presented here to calculate the average genetic distance between all viruses, living and final, is suitable for compartmental models in general\cite{murray2007mathematical}. Although we develop it to the SEIR model, it can be applied to other models of this type. From now on we shall abbreviate {\it average genetic distance} by {\it average distance} for simplicity.

\subsubsection*{Single initial infection}

{ Here we assume that the epidemic starts with a single infected individual.} Our goal is to compute the average distance $d_{t+1}$ at time $t+1$ given the average distance $d_t$ at time $t$. Notice that at the \textit{beginning of iteration} $t+1$, there are different \textit{kinds} of viruses: those that are \textit{already final} and have ceased to evolve (whose number is $R_t$); \textit{viruses hosted} in exposed individuals ($E_t$), thus still evolving; and also those \textit{hosted in infected} individuals ($I_t$). During the iteration, \textit{new infections} appear ($x_t$) and some \textit{infected individuals recover} ($r_t$), and thus do not evolve at this time step. Then, given $d_t$, we calculate the new average distance \textit{between each kind of virus} which exists at the \textit{end of iteration} $t+1$, as well as the new average distance \textit{within each kind of virus}. 

Given that $\mu\ll1$, we consider that the probability that two mutations happen in the same nucleotide in the course of the epidemic is negligible. This is a good approximation if the epidemic duration $T$ remains sufficiently small, $\mu T\ll1$. We also consider that each new infection in the same iteration comes from different hosts, which is valid for $R_0/\tau_0<1$, with $\tau_0$ the average duration of symptoms. This means that we do not expect more than one new infection per infected individual in a single iteration. Highly connected nodes, however, can break this assumption, giving rise to super-spreaders. Network heterogeneity, therefore, can show deviations from our estimation. Under these assumptions, the new average distance  (at the end of iteration $t+1$) among the $E_t$ is $d_t+2B\mu$, once they distanced $d_t$ at the begging of iteration $t+1$ and evolved along the iteration, each virus getting $B\mu$ mutations. The new average distance between the $E_t$ and the $R_t$ is $d_t+B\mu$, since only the $E_t$ evolved. We emphasize that the approximations used in this section are only for simplification of the analytical equations; the simulations in Section \nameref{sec:secResults} run as previously described. 



Once all average pairwise distances have been calculated, $d_{t+1}$ is given by a weighted average, where the weigths are the number of pairs sharing that distance. For instance, the number of pairs between exposed and recovered individuals is $E_t  R_t$, while the number of pairs within exposed individuals is $E_t(E_t-1)/2$.
 
%

All distances are calculated in \nameref{appendixes}, and we find the recurrence equation
\begin{align}
    d_{t+1}=\frac{1}{Z_t}&\left(d_t(R_t+E_t+I_t)(R_t+E_t+I_t-1)\right.\nonumber\\+&x_td_t\left(1+2B\mu\frac{R_t}{I_t+E_t+R_t}\right)(x_t-3+2R_t+2I_t+2E_t)\nonumber\\+&\left.2B\mu(E_t+I_t-r_t)(E_t+I_t+R_t+x_t-1)\right)\label{eq:recorrencia}
\end{align}
where $Z_t=(R_t+E_t+I_t+x_t)(R_t+E_t+I_t+x_t-1)$, $r_t=R_{t+1}-R_t$ and $x_t=(E_{t+1}-E_t)+(I_{t+1}-I_t)+(R_{t+1}-R_t)$ . 
%

Therefore, given the epidemic curves $S_t$, $E_t$, $I_t$ and $R_t$,  respectively the Susceptible, Exposed, Infected and Recovered at time $t$, we can infer the evolution of average genetic distances. Taking the limit of continuous time between events we find the approximation,
\begin{align}
    \dot{d}=\frac{2\dot{S}d\left(1-B\mu R\left(2-\frac{3}{N-S}\right)\right)}{(N-S)(N-1-S)}+2B\mu\left(1-\frac{R}{N-S}\right)\label{eq:distmeanfield}
\end{align}
where $N-S=I+R+E$ and $\dot{S}=-(\dot{E}+\dot{I}+\dot{R})$. The derivation of this limit are described in \nameref{appendixes}. Since this equation depends only on the continuous curves $S(t)$ and $R(t)$, the initial and final compartment, it can be added to the classic SEIR model to infer the genetic evolution, or to the SIR model, if the exposed compartment is kept empty, meaning that all hosts are infectious.  This result holds if viral evolution occurs in the same way in every intermediate compartment and if every virus passes through all compartments. Adding more compartments with different dynamical behavior or changing the mutation mechanism through different compartments would change the equations (\ref{eq:recorrencia}) and (\ref{eq:distmeanfield}) but the procedure described in the begging of this section to find $d_{t+1}$ should remain the same.

\subsubsection*{\label{subsec:many}Multiple initial infections}

Eq.(\ref{eq:recorrencia}) considers the epidemic starting with a single infected individual. To consider $m>1$ initial infections, we must include the distance among the $m$ different lineages. Let $\mathfrak{D}_t$ be the average distance among all viruses at time $t$, $d_t^{(i)}$ the average distance among the viruses of lineage $i$ at time $t$, $d_0^{(ij)}$ the distance between the initial viruses $i$ and $j$, and $d_{root,t}^{(i)}$ the average distance at time $t$ of lineage $i$ to the root of lineage $i$. Thus,
\begin{align}
    \mathfrak{D}_t=&\left[\sum_{i=1}^{m}d_t^{(i)}\left(R_t^{(i)}+E_t^{(i)}+I_t^{(i)}\right)\left(R_t^{(i)}+E_t^{(i)}+I_t^{(i)}-1\right)/2\right.\nonumber\\
    &+\left.\sum_{i=1}^{m-1}\sum_{j=i+1}^{m}\left(d_0^{(ij)}+d_{root,t}^{(i)}+d_{root,t}^{(j)}\right)\left(R_t^{(i)}+E_t^{(i)}+I_t^{(i)}\right)\left(R_t^{(j)}+E_t^{(j)}+I_t^{(j)}\right)\right]\nonumber\\
    \div&\left[\left(\sum_{i=1}^{m}\left(R_t^{(i)}+E_t^{(i)}+I_t^{(i)}\right)\right)\left(\sum_{i=1}^{m}\left(R_t^{(i)}+E_t^{(i)}+I_t^{(i)}\right)-1\right)/2\right]\label{eq:rec3}
\end{align}
where $R_t^{(i)}$, $E_t^{(i)}$ and $I_t^{(i)}$ are, respectively, the number of recovered, exposed and infected individuals of lineage $i$ at time $t$.
The first sum represents the distances within each lineage $i$, while the double sum is due to the distance between each pair of lineages $i$ and $j$.
In this equation, we assume the $\mu\ll1$ (for coronaviruses, $\mu$ lies in the range $\sim[10^{-5},10^{-2}]$ per site per year\cite{zhao2004moderate}) so that mutations for each virus are unlikely to occur twice at the same nucleotide.

For each lineage $i$, $d_t^{(i)}$ can be calculated from Eq.(\ref{eq:recorrencia}) or Eq.(\ref{eq:distmeanfield}) and $d_0^{(ij)}$ must be a given matrix. The distance $d_{root,t}^{(i)}$ can be calculated similarly as Eq.(\ref{eq:recorrencia}),
\begin{equation}
    d_{root,t+1}^{(i)}=d_{root,t}^{(i)}+\frac{B\mu}{E_t^{(i)}+I_t^{(i)}+R_t^{(i)}+x_t^{(i)}}\left(E_t^{(i)}+I_t^{(i)}-r_t^{(i)}+\frac{4x_t^{(i)}R_t^{(i)}d_{root,t}^{(i)}}{E_t^{(i)}+I_t^{(i)}+R_t^{(i)}}\right)
    \label{eq:rec2}
\end{equation}
with the continuum limit
\begin{equation}
    \dot{d}_{root}=B\mu\left[1-\frac{R^{(i)}}{R^{(i)}+I^{(i)}+E^{(i)}}\left(1-\frac{4d_{root}(\dot{E}^{(i)}+\dot{I}^{(i)}+\dot{R}^{(i)})}{R^{(i)}+I^{(i)}+E^{(i)}}\right)\right]
    \label{eq:meanfield2}
\end{equation}
where $R^{(i)}$, $I^{(i)}$ and $E^{(i)}$ are SEIR variables for lineage $(i)$. The details behind these results are described in \nameref{appendixes}. 

\subsection*{ Viral spread throughout communities}

{

As an application of { our model} and computational framework, we studied the genetic evolution of a viral spread throughout four weakly and linearly connected communities, i.e., a network with four modules, representing different cities. The goal is to understand how the average genetic distance between viruses in distant communities change if the connectivity between the intermediary communities changes.

We start by generating four independent Barabasi-Albert networks, named 1, 2, 3 and 4. Then, we connect individuals from networks $i$ and $i+1$ with probability $p$ in a way they form a line of communities. The Barabasi-Albert network is chosen in order to include heterogeneity in the contact network \cite{MARQUIONI2020109999}. Finally, we analyse the average genetic distance between viruses from cities 1 and 4 for different values of $p$. The epidemic starts with a single infected individual in city 1 and spreads through the entire network.

Although in our model we always consider that individuals acquire perfect cross-immunity against all strains after being infected the cross-immunity could { in principle} be lost if a new infecting virus were { too different from the original infection}. Thus, if the distance between viruses from cities 1 and 4 is large, an infected individual from city 4 that travels to city 1 might reinfect an already recovered individual. { Although our simulations do not include this possibility,}  this is an interesting way to investigate how the risk of reinfection changes due to changes in the network topology.


}

\section*{Results and Discussion\label{sec:secResults}}

\subsection*{Single initial infection}

\begin{figure}
\includegraphics[width=14cm]{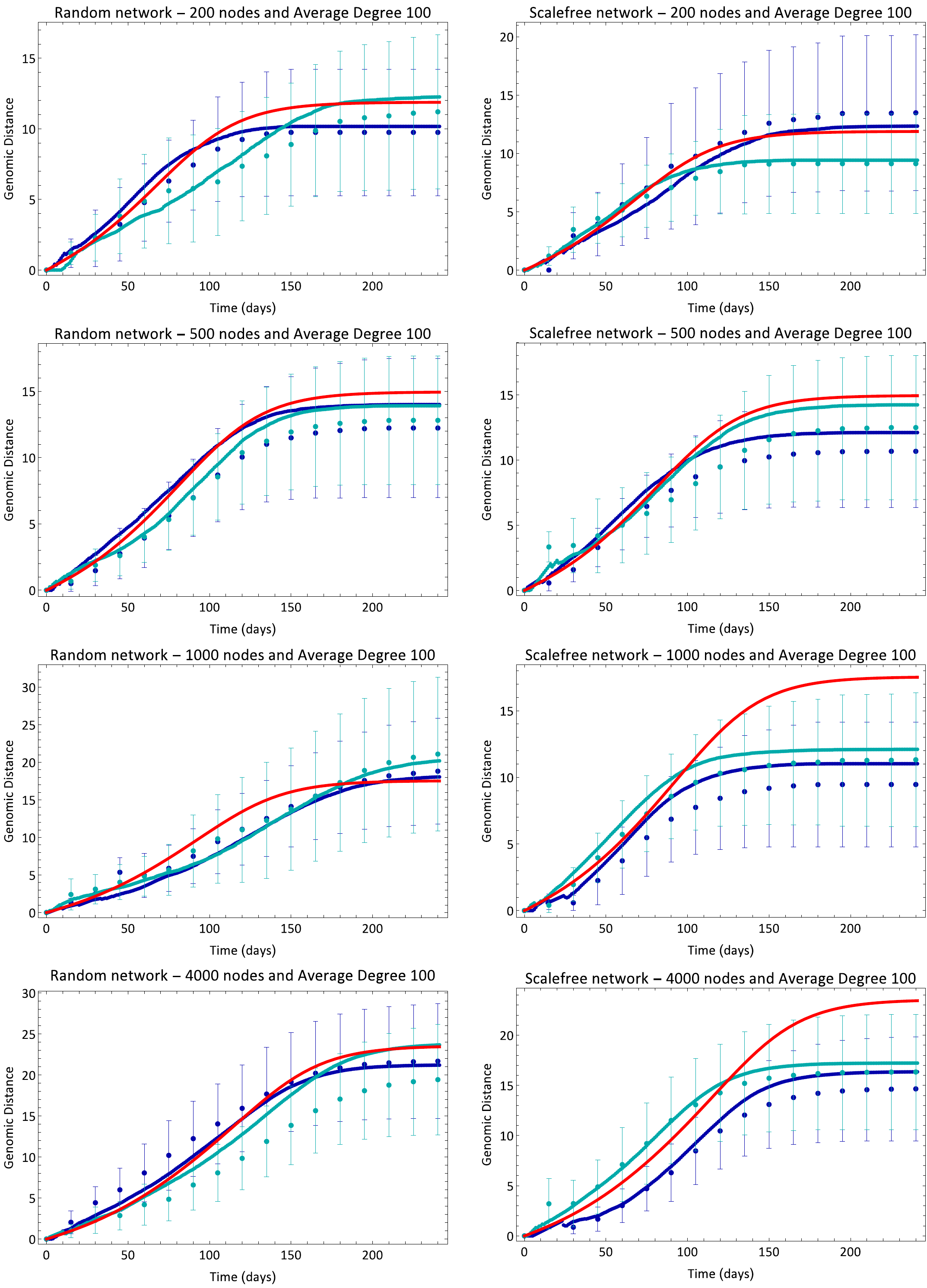}
\caption{{\bf Evolution of average genetic distance.} Blue lines and dots are, respectively, analytical (Eq.(\ref{eq:recorrencia})) and simulation results for different simulations. Different shades of blue correspond to different simulations for the same set of parameters. The red line shows the result of mean-field Eq.(\ref{eq:distmeanfield}). Error bars are standard deviation of the distance distribution in each simulation at each time.}
\label{fig:AllDists}
\end{figure}

We ran our model for random (Erdos-Renyi) and scalefree (Barabasi-Albert) networks and calculated the average genetic distance. We used networks of 200, 500, 1000 and 4000 nodes, and average degree of 100 nodes. The infection starts with a single infected individual chosen at random and evolves according to { the description in section 2} \ref{sec:model}. Fig.\ref{fig:AllDists} shows comparisons between the simulated distance and the average distance calculated from Eq.(\ref{eq:recorrencia}) and Eq.(\ref{eq:distmeanfield}). Each subfigure contains two different simulations and the mean-field solution for that respective set of parameters. We see that that Eq.(\ref{eq:distmeanfield}) approaches Eq.(\ref{eq:recorrencia}) only for Erdos-Renyi networks, since only this topology mimics the well-mixed hypothesis considered in mean-field models. Because each genetic evolution curve is calculated from the corresponding epidemic curves, we cannot average over many simulations, thus the error bars are simply the standard deviation of the distribution of distances among all viruses that appeared at that specific simulation time step. Another important feature of this analytical formulation is that, once it is an average description, it does not capture the random appearance or extinction of viral lineages, which can introduce important deviations from our analytical description. 

\subsection*{Multiple initial infections}

\begin{figure}
\includegraphics[width=14cm]{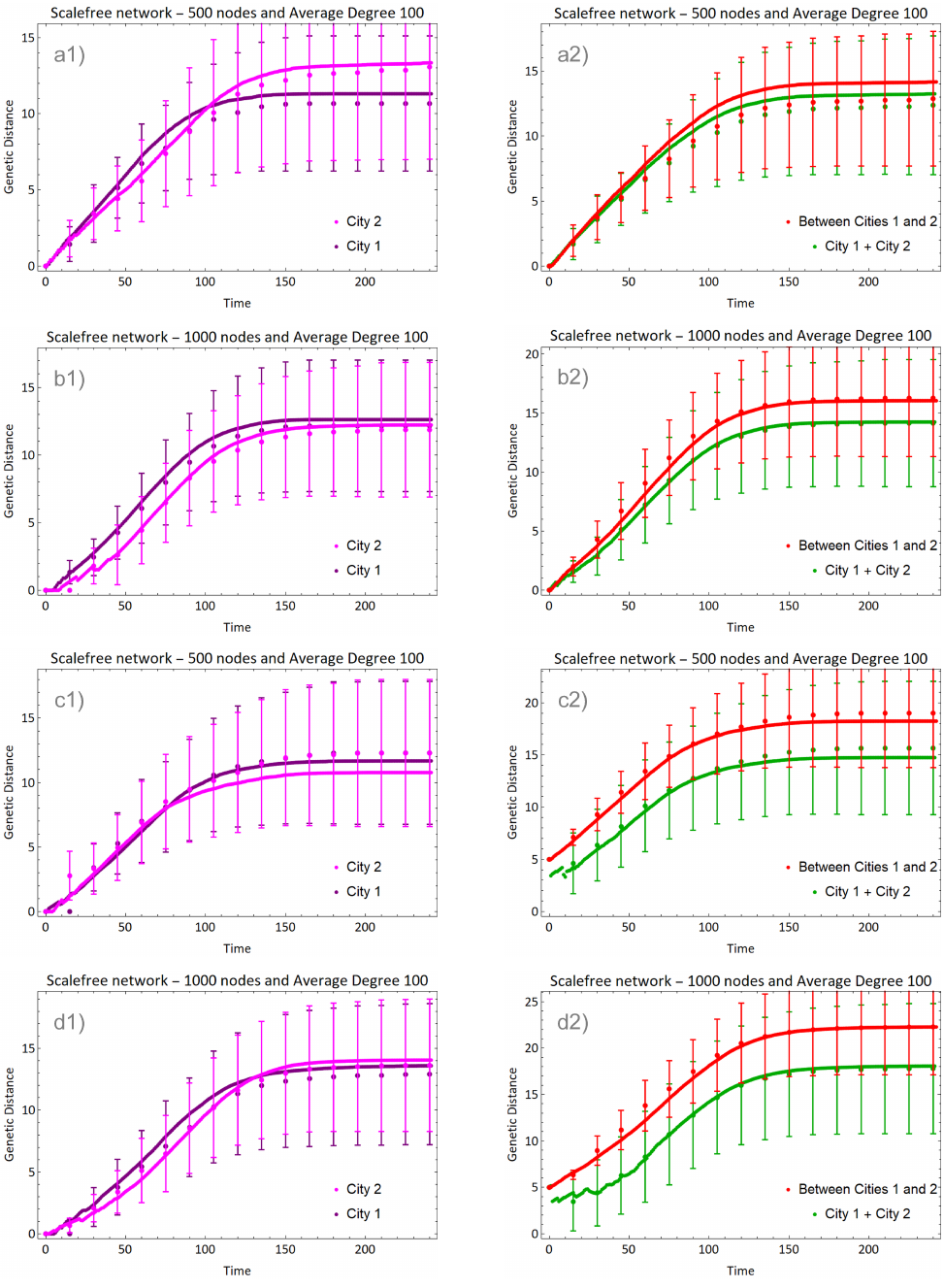}
\caption{{\bf Evolution of average genetic distance in two isolated cities (sizes indicated in the panels).} In (a) and (b) the initial viruses were identical and in (c) and (d) they differed by 5 nucleotides. Lines show the average distance within each city (pink), between cities (red) and total average distance (green).}
\label{fig:AllDists2}
\end{figure}

Fig.\ref{fig:AllDists2} shows the evolution of epidemic in two different cities  (non-connected networks of random and scalefree types), each one starting its infection with a single infected individual chosen at random. The evolution in each city is calculated with Eq.(\ref{eq:recorrencia}) (pink curves), while the distance between cities 1 and 2 is $d_t^{(1,2)}=d_0^{(1,2)}+d_{root,t}^{(1)}+d_{root,t}^{(2)}$, where $d_{root,t}^{(i)}$ is calculated with Eq.(\ref{eq:rec2}) (red curve) and the total average distance $\mathfrak{D}_t$ (green curve) is given by Eq.(\ref{eq:rec3}). The initial distance between the viruses that infected each city is $d_0^{(1,2)}=0$ in panels (a) and (b), and $d_0^{(1,2)}=5$ in panels (c) and (d).

\subsection*{\label{subsec:china}  The COVID-19 epidemic in China}

Eq.(\ref{eq:recorrencia}) describes the evolution of average genetic distance between viruses in a single community and depends only on the epidemic curves. It might, therefore, be used to estimate the genetic evolution in real cases. The { beginning of} COVID-19 epidemic in China is a suitable example, considering the existence of a single patient zero. In any other country, the epidemic may have started with more than one individual, which would require the difficult task of tracking the lineages. { The same applies to secondary waves of infection in China.} 

We obtained Chinese data from the Wolfram Data Repository\cite{wolfram2020epidemic}, and corrected it as in reference \cite{ivorra2020mathematical}. Because of the existence of undetected cases, we estimated the real number of cases considering references \cite{ivorra2020mathematical,li2020substantial}. Because the number of exposed individuals is not directly available we choose to consider the simpler SIR model in this case. Notwithstanding, because the cases notification started only in January while the epidemic started in December, we extrapolated the data to previous dates, in order to calculate the genetic evolution since patient zero, as we have made in Fig.\ref{fig:AllDists}. All these data corrections and considerations are described in the supporting information.

To compare the result of Eq.(\ref{eq:recorrencia}) with the real genetic evolution, we used {  carefully selected} 55 real genomes sequenced and collected in China, also available in the Wolfram Data Repository\cite{wolfram2020genomas}. The Hamming distance between each pair of genome was obtained by first aligning every two genomes with the Needleman-Wunsch algorithm with score matrix $+1$ for match and $-1$ for mismatch\cite{sung2009algorithms}. Then, we considered the Hamming distance between a given pair of genomes as the number of mismatches that are not {\it indels}, i.e., we considered only nucleotide substitutions. The algorithm to estimate the distance evolution is explained in \nameref{appendixes}, as we also detail the informations of the used genetic data.

\begin{figure}[H]
\includegraphics[width=14cm]{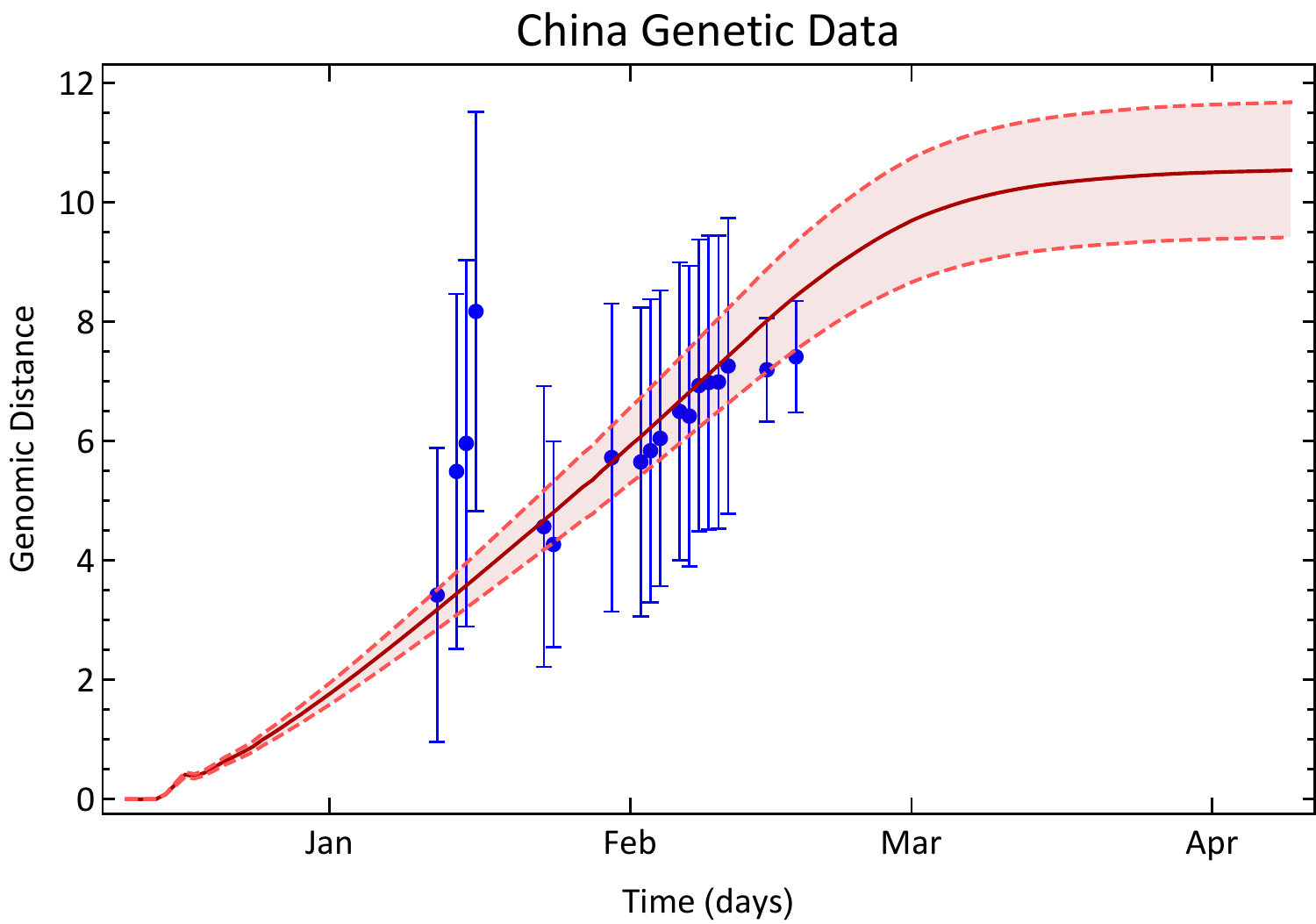}
\caption{{\bf   The genetic evolution of SARS-CoV-2 in China.} Blue dots are the genetic distance among SARS-CoV-2 inferred from data {  collected in China between 12/23/2019 and 03/24/2020}. {  The error bars are standard deviation of pairwise distance propagated through the equations.} The brown line shows the genetic distance estimated with Eq.(\ref{eq:recorrencia}) and the Chinese epidemic data. The interval around the brown curve is a $\pm10\%$ error interval on the value $B\mu$, which we considered to be $B\mu=29900\times0.001/365$.}
\label{fig:China}
\end{figure}

Fig.\ref{fig:China} shows the result obtained from Eq.(\ref{eq:recorrencia}) (brown line) and the estimated genetic evolution (blue dots). The interval around the brown line is an error of $\pm10\%$ on the product $\mu B$, which is the only parameter in the equation (\ref{eq:recorrencia}). Despite all corrections to the epidemic data and the small number of real genomes we used to infer the real genetic evolution, except for a few points, all the inferred average genetic distances between RNA sequences lie in the predicted interval given by our theoretical model. {  Because the epidemic in China was readily contained, the average distance $d_t$ saturated. }

\subsection*{\label{sec:results2}Communities and reinfection}

\begin{figure}[!h]
\includegraphics[width=14cm]{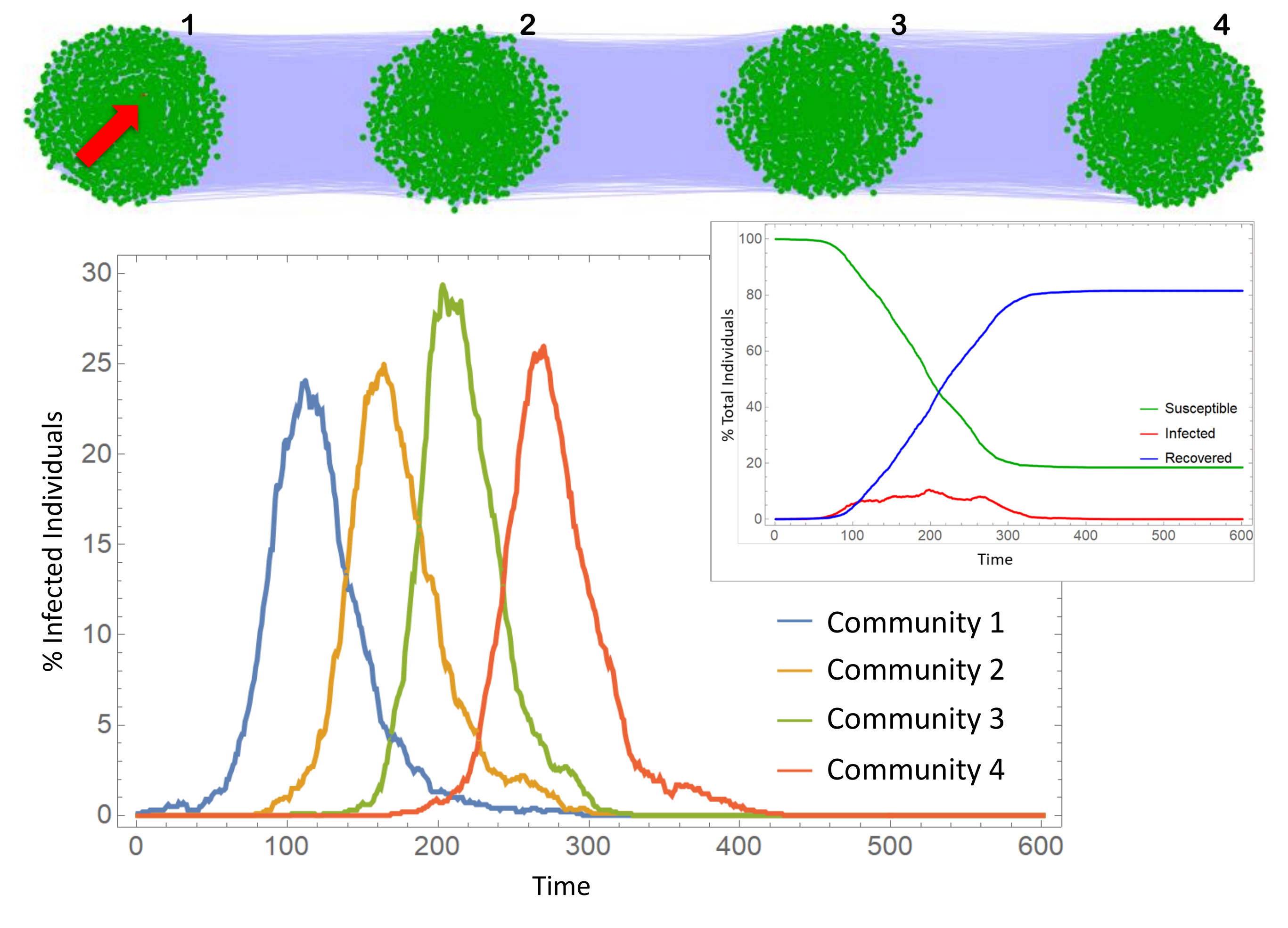}
\caption{{\bf Contact network of four communities on a line and infection curves.} Communities are Barabasi-Albert networks with 1000 nodes. The infection starts with a single infected individual in the first community (red node indicated with the red arrow). The epidemic parameters are in \nameref{appendixes}}.
\label{fig:linha}
\end{figure}

In this section, we  consider the spread of the epidemic through four communities, representing cities, connected linearly as in Fig.\ref{fig:linha}.  
Fig.\ref{fig:linha} shows an example of the contact network. From left to right, we number the communities, or cities, from 1 to 4. The epidemic starts with a single infection in city 1 and spread through the entire network. Fig.\ref{fig:linha} also shows the Infection curves obtained from a simulation. The infection peak delay from one city to other is responsible for the plateau-type curve of total infections.

\begin{figure}[!h]
\includegraphics[width=14cm]{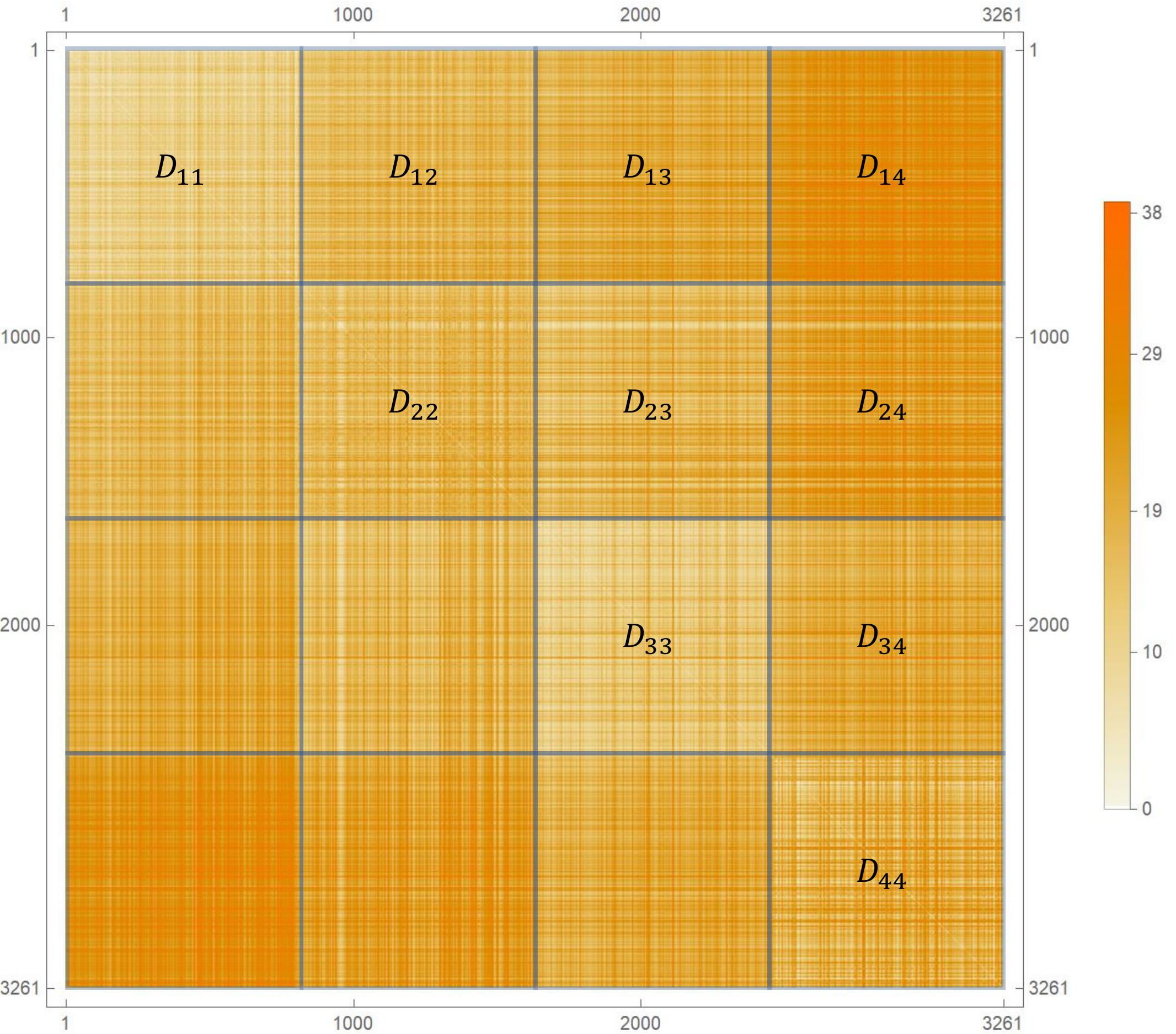}
\caption{{\bf Hamming distance between pairs of viruses.} The distance matrix is sorted by the city. Diagonal blocks show the distance between the viruses from a single city, while the non-diagonal blocks are the distances between the viruses from different cities.}
\label{fig:hammingDist}
\end{figure}

To analyse the genetic evolution in this system we simulated the dynamic until the epidemic was over and calculated the Hamming distance between every pair of final genomes $\alpha$ and $\beta$, constructing the distance matrix $d^{\alpha\beta}$ (Fig.\ref{fig:hammingDist}). Viruses are ordered according to their position in the line, i.e., first the genomes from city 1, then those from the city 2, and so on. We calculated the average distances $D_{i-j}$ between the final genomes from cities $i$ and $j$ and compared with $D_{i-i}$, the average distance within city $i$. 

\begin{figure}[!h]
\includegraphics[width=14cm]{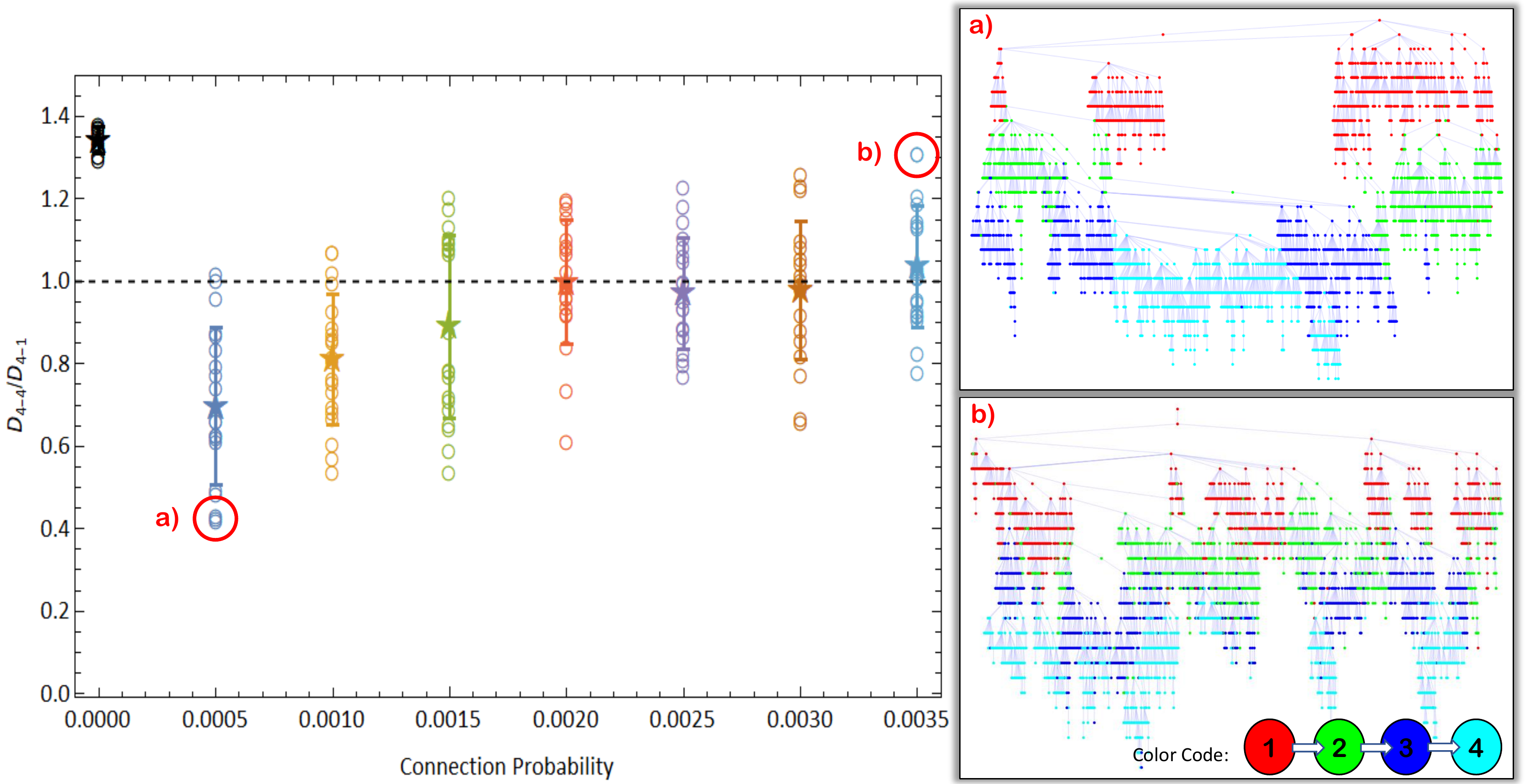}
\caption{{\bf Ratio between the average distance in city 4 and the average distance between cities 1 and 4.} Right panels show infection trees for the simulations highlighted with red circles. Open circles show results for individual simulations, the star is the average over 20 simulations and error bars are standard deviations. $p=0$ represents a single Barabasi-Albert network with 4000 nodes (see text). Nodes in infection trees represent infected individuals, colored according to its city. City 4 (cyan) in panel (a), where $D_{4-4}/D_{4-1}<1$, was almost entirely infected by a single viral lineage, while in panel (b) where $D_{4-4}/D_{4-1}>1$, it was infected by many different viral lineages.}
\label{fig:D44D41}
\end{figure}

\begin{figure}[!h]
\includegraphics[width=14cm]{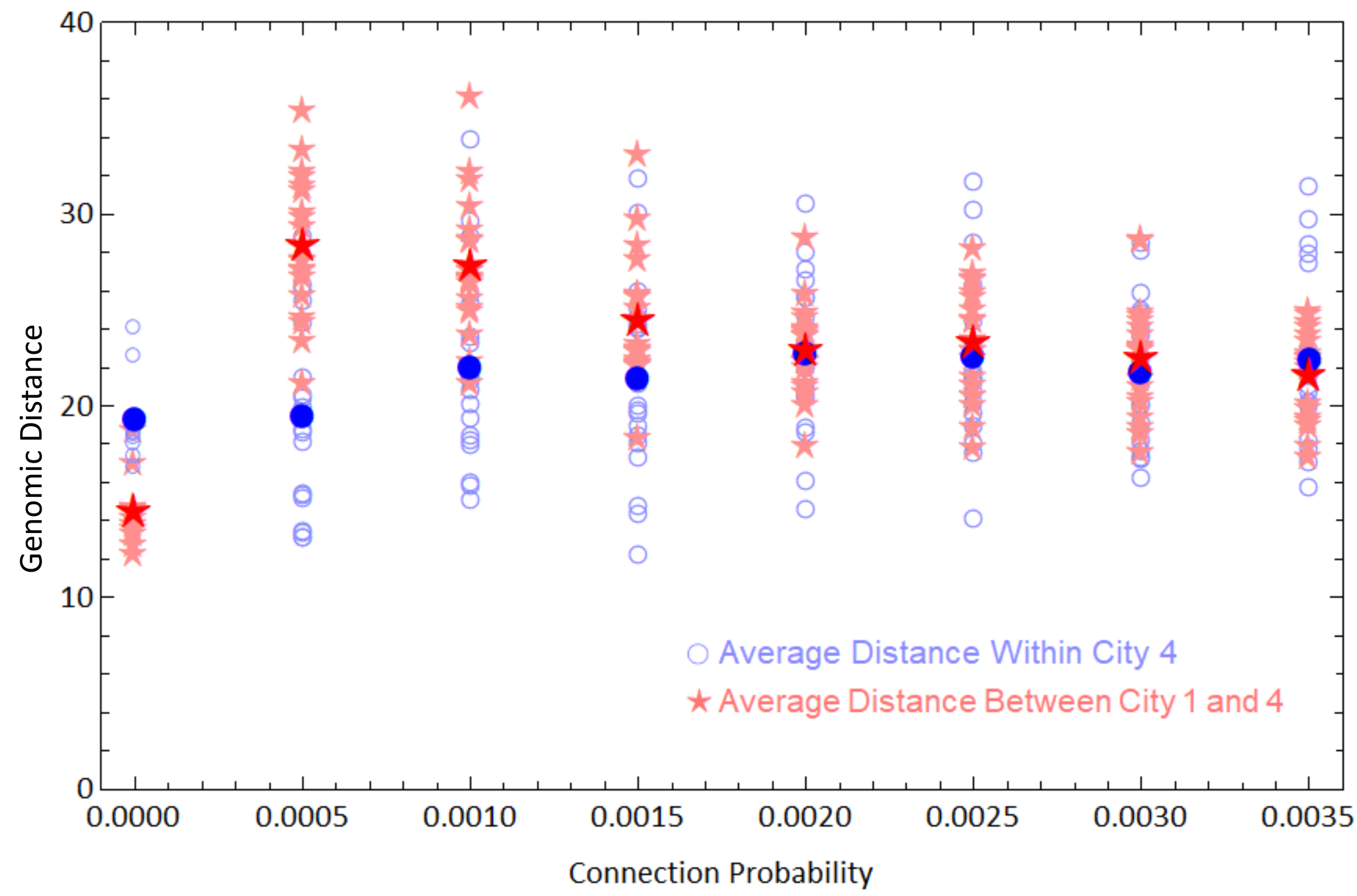}
\caption{{\bf Average genetic distances within cities 1 and 4.} Open blue circles are average distance between the viruses of city 4 from a single simulation, and the filled blue circle is average of these values. Light red stars are average distances between viruses from cities 1 and 4 and the dark red star is the average of these values. We ran 20 simulations for each value of connection probability.}
\label{fig:reinfec}
\end{figure}

{  As a null model, we run the epidemic over a single Barabasi-Albert network wih the total size of the 4 cities. City $i$, in this case, means the i-th quarter of the infected nodes. We plot the results of the null model as $p=0$ in Fig.\ref{fig:D44D41} and Fig.\ref{fig:reinfec} for comparison. The single network behaves very differently from the four module network, not showing the same interesting results we find for the communities.}

Fig.\ref{fig:D44D41} shows the ratio $D_{4-4}/D_{4-1}$ as a function of the connection probability $p$. The results are averages over 20 different simulations for 7 different values of $p$. When $p$ is small, $D_{4-4}/D_{4-1}<1$, meaning that the viruses from city 4 are, in average, closer to each other than they are to the viruses from city 1. When $p$ increases, the ratio $D_{4-4}/D_{4-1}$ approaches $1$, indicating that the viruses from city 4 are so close to each other as they are to viruses from city 1.

In order to understand the origin of this effect we analyse the infection trees in each case (Fig.\ref{fig:D44D41}, left). Each node in the trees represents a recovered individual and is connected upwards with whoever infected it. Colors represent cities and it is possible to count how many initial infections each city had along the epidemic, i.e., how many lineages has infected each city. When $p$ is small, very few lineages were responsible for infecting city 4 but for higher values of $p$, this number increases. { This is expected, since more connected communities should have more infection gates.} This result is a consequence of the founder effect, i.e., only a few individuals, ``the founders", give rise to a new population in the new location\cite{forster2020phylogenetic,ruan2020founder}.  However, the system passes through a non-trivial bistable point. When $p=0.0015$, the values of $D_{4-4}/D_{4-1}$ accumulate around two different values, one above $1$ and another below $1$. In this case the average is not a good descriptor of the actual system behaviour and there is a competition between different lineages infecting city 4. In simulations where $D_{4-4}/D_{4-1}>1$, many lineages were successful in infecting the city 4, whereas when $D_{4-4}/D_{4-1}<1$, only a few did so successfully.

Fig.\ref{fig:reinfec} shows the values $D_{4-4}$ and $D_{4-1}$ obtained in each simulation. The average over simulations of the average distance within the forth city $D_{4-4}$ (highlighted blue circles) does not change considerably with $p$ (around $D\approx21$ nucleotides). Under a neutral evolutionary perspective, viruses will belong to different strains if they differ by more than $G$ nucleotides, where $G$ is a parameter {  whose value depends on the virus} \cite{costa2018registering,de2017speciation}. If $D>G$, viruses in city 4 would belong, on average, to different strains when compared to city 1. As an example, if $G=26$ new strains would arise, on average, in city 4 for $0<p \leq 0.0010$, allowing a recovered individual from city 1 to be reinfected by an infected individual from city 4 if they are put in contact with each other (by travelling, for instance). Therefore, there is an increased risk of reinfection due to low connectivity among communities. In this sense, pandemics are more likely to originate new strains than epidemics, as they affect far more distant (therefore less connected) communities. One confirmed case of reinfection by COVID-19 in Hong-Kong had the virus differing by 24 nucleotides from the first infecting virus\cite{to2020covid}. This distance matches a value for $G$  {  for which the network connectivity would strongly influence the rise of reinfections.}

\section*{\label{sec:discussion}Conclusions}

We have introduced an individual based model to describe the genetic evolution of a RNA-virus epidemic spreading . We used the SEIR model { with four compartments} on networks, but the evolutionary dynamics can be implemented in more compartmentalized epidemic models. 
We provided an analytical description that can be generalized for models with more compartments. An important result of this study is the mean-field approximation,  Eq.(\ref{eq:distmeanfield}), for the evolution of the average genetic distance, which can be added directly to the mean-field { SIR or} SEIR models.

Our analytical description of the average genetic distance between viruses is neutral and depends only on the epidemic curves. This allows us to project the evolutionary scenario without using the actual genome sequences. Deviations from these predictions in genetic data could reveal the strength of selection or network effects.  We compared our prediction using only fifty complete genomes sequenced and collected in China and found good agreement. 

We { have} also analysed the genetic evolution of the epidemic when it spreads over different communities. By changing the connection probability $p$ between 4 linearly arranged communities we investigated how different the viruses infecting city 4 would be from their ancestors in city 1. Our simulations showed that when $p$ is sufficiently small, the genetic difference between these viruses can be quite large, spanning 30 loci. This could allow an infected individual from city 4 to reinfect a recovered individual from city 1. This is a consequence of the founder's effect, which is stronger if $p$ is small as it decreases the number of infection gates of a community. Therefore, we expect increased risk of reinfection from contacts between travelling individuals living in distant territories.

{ Although the computational framework we described for the viral evolution is neutral, it can be adapted to including other evolutionary aspects, such as differential fitness for mutations in certain genome regions or loss of cross-immunity. These and other features are important topics to be added and studied in future works.}

\section*{Supporting information}


\paragraph*{S1 Appendix}
\label{appendixes}
Simulation parameters, analytical calculations, real genetic evolution algorithm and Chinese epidemic data corrections.

\paragraph*{S1 Table}
\label{s1:table}
{\bf All Chinese genome sequences.} All genomes registered in Wolfram Repository ``Genetic Sequences for the SARS-CoV-2 Coronavirus'' with complete \emph{NucleotideStatus} and human \emph{Host} from China (data accessed 19/08/2020).

\paragraph*{S2 Table}
\label{s2:table}
{\bf Included sequences sorted by Collection Date.} All informations according to S1 Table.

\paragraph*{S3 Table}
\label{s3:table}
{\bf Genome information used to calculate points in Fig.\ref{fig:China}.} We have used a 14 days time window, i. e., every sequenced genome within an interval of 14 days were considered as infected ones, while the previous were considered to be recovered.

\section*{Acknowledgments}
We thank Dr. Débora Pricepe,  Dr. Flávia D. Marquitti and Luis F.P.P.F. Salles for critical readings and valuable suggestions. This work was supported by the S\~ao Paulo Research Foundation (FAPESP), grants 2019/13341-7 (VMM), 2019/20271-5 and 2016/01343-7 (MAMA), and by Conselho Nacional de Desenvolvimento Cient\'ifico e Tecnol\'ogico (CNPq), grant 301082/2019-7 (MAMA). 


%
%
%


\providecommand{\noopsort}[1]{}\providecommand{\singleletter}[1]{#1}%



\section*{Supplementary material}

# Modeling viral mutations in the spread of epidemics - Appendixes

Vitor M. Marquioni[1], Marcus A. M. de Aguiar[1*]

**1** Instituto de Física "Gleb Wataghin", Universidade Estadual de Campinas - UNICAMP, Campinas, SP, Brazil

* aguiar@ifi.unicamp.br

## 1 Appendix A: Simulation parameters

The network simulations follow the model proposed in reference [1], and the parameters are displayed in Table 1.

| Parameter | Value |
|---|---|
| $R_0$ | 2.4 [2] |
| Average Symptoms Duration $\tau_0$ | 14 days [3,4] |
| Networks Average Degree $D$ * | 100 [1] |
| Incubation Time Distribution $\mathcal{P}(\tau)$ | $\Gamma(6.25, 25/26)$ [5] |
| Mutation Rate $\mu$ | 0.001 substitution per base, per year [6,7] |
| Genome Size $B$ | 29900 bases [2] |

**Table 1. Simulation Parameters.** The number of nodes in each simulation is described properly.
*This is the input average degree for the networks construction, but the actual value for each realization fluctuates. For the communities simulations, this is the parameter for constructing each isolated network, as also for the control case $p = 0$.

For the numerical solution of mean-field approaches, following the SEIR model

$$\begin{aligned}\dot{S} &= -\beta SI/N \\ \dot{E} &= \beta SI/N - \sigma E \\ \dot{I} &= \sigma E - \gamma I \\ \dot{R} &= \gamma I\end{aligned} \tag{A1}$$

we have used the following parameters: $R_0 = 2.4$, $\gamma = 1/14$ day$^{-1}$; $\beta = R_0 \gamma$ and $\sigma = 1/\langle t_i \rangle$, where $\langle t_i \rangle$ is the mean period of incubation, averaged over the distribution from Table 1 [1].

## 2 Appendix B: Analytical calculations

Our goal is to derive a recurrence equation for the average genetic distance, i.e., given the distance $d_t$ at time $t$, we aim to calculate the distance $d_{t+1}$ at time $t+1$. The idea is to calculate $d_{t+1}$ as a weighted average, where the weights are the number of pairs that are distanced by a certain amount. In a SEIR model, every iteration starts with a given number of recovered ($R_t$), infected ($I_t$) and exposed ($E_t$) individuals. When an individual recovers, its infecting virus stops to spread and to evolve, and we call it a

*final virus*. There are $R_t$ final viruses at the beginning of a given iteration. Viruses infecting Exposed individuals can mutate during this iteration. However, viruses in Infected individuals can either evolve and mutate in this time step or not, since their hosts might recover. The latter become final and are counted as $r_t$. Infected individuals can also spread the virus, which replicate before evolving or becoming final. Such *offspring* ($x_t$) increase the number of viruses in Exposed individuals in the next iteration, when they will be allowed to evolve.

At the beginning of iteration $t+1$, there are $(R_t + E_t + I_t)(R_t + E_t + I_t - 1)/2$ pairs of viruses sharing an average distance equal to $d_t$, but along the iteration some of the distances may increase by a certain amount to be calculated, as also new viruses may arise. Therefore,

$$d_{t+1} = \frac{1}{Z'_t}\left(d_t \frac{(R_t + E_t + I_t)(R_t + E_t + I_t - 1)}{2} + \text{Increases} + \text{Offspring}\right), \quad \text{(A2)}$$

where $Z'_t$ is a normalization factor, which counts the total number of pairs at the end of iteration $t+1$,

$$Z'_t = \frac{(R_t + E_t + I_t + x_t)(R_t + E_t + I_t + x_t - 1)}{2}. \quad \text{(A3)}$$

If the mutation rate is zero and no new infections occur ($x_t = 0$) the "Increases" term and the "Offspring" term are equal to zero, and $d_{t+1} = d_t$, as expected.

In the following two subsections, we shall calculate the "Increases" term and the "Offspring" term, which accounts for the evolution and for the spread, respectively.

## 2.1 Increases

Genetic distances between evolving viruses increase over time. In order to calculate how much these distances increase we first consider that mutations occurring in the same locus of different genomes are unlikely, as well as more than one mutation per locus on a single genome. This approximation holds as long as the epidemic duration $T$ remains sufficiently small, $\mu T \ll 1$. Thus, after one time step, an evolving genome acquires, on average, $B\mu$ mutations. The distance between two evolving genomes will increase, on average, by $2B\mu$ nucleotides after one time step. The distance between viruses in exposed individuals, for example, increases by $2B\mu$ and because there are $E_t(E_t - 1)/2$ pairs of exposed individuals, their evolution along the iteration $t+1$ contributes $2B\mu E_t(E_t - 1)/2$ to the Increases term. On the other hand, the distance between viruses in an exposed and a recovered individual, or an infected individual that recovers, is only $B\mu$, because the latter two do not evolve. There are $E_t(R_t + r_t)$ pairs among these viruses, and thus their contribution to Increases is $E_t(R_t + r_t)B\mu$. We recall that the updates in our model occur in the order "Transmission", "Attempt to Recovery" and lastly, "Genome Evolution". Thus, if an infected individual recovers its virus does not have the chance to mutate.

Therefore, in order to compute the Increases term, we must calculate the average increase in distance between all pairs of viruses and how many pairs of these viruses exist. Table 2 summarizes this information. We obtain

$$\begin{aligned}\text{Increases} =\;& E_t R_t B\mu + E_t r_t B\mu + (I_t - r_t)r_t B\mu + (I_t - r_t)R_t B\mu + (I_t - r_t)E_t 2B\mu \\ & + \frac{E_t(E_t - 1)}{2} 2B\mu + \frac{(I_t - r_t)(I_t - r_t - 1)}{2} 2B\mu. \end{aligned} \quad \text{(A4)}$$

Table 2. Increases in average distance and number of pairs of viruses.

| Viruses | Number of Pairs | Average Distance Increase |
|---|---|---|
| $(E_t)$ and $(R_t)$ | $E_t R_t$ | $B\mu$ |
| $(E_t)$ and $(r_t)$ | $E_t r_t$ | $B\mu$ |
| $(I_t - r_t)$ and $(r_t)$ | $(I_t - r_t) r_t$ | $B\mu$ |
| $(I_t - r_t)$ and $(R_t)$ | $(I_t - r_t) R_t$ | $B\mu$ |
| $(I_t - r_t)$ and $(E_t)$ | $(I_t - r_t) E_t$ | $2B\mu$ |
| $(E_t)$ and $(E_t)$ | $E_t(E_t - 1)/2$ | $2B\mu$ |
| $(I_t - r_t)$ and $(I_t - r_t)$ | $(I_t - r_t)(I_t - r_t - 1)/2$ | $2B\mu$ |
| $(R_t)$ and $(R_t)$ | $R_t(R_t - 1)/2$ | 0 |
| $(r_t)$ and $(r_t)$ | $r_t(r_t - 1)/2$ | 0 |
| $(r_t)$ and $(R_t)$ | $r_t R_t$ | 0 |

## 2.2 Offspring

The contribution of the new infections to the average distance $d_{t+1}$, the Offspring term, is more tricky. To simplify matters we will assume that an infected individual infects only one susceptible per time step, which is a good assumption if the basic reproduction number $R_0$ is small compared to the average duration of symptoms. Thus, $x_t$ is also the number of individuals who infected a susceptible within the time step $t+1$, which will be called *ancestors* from now on. Let $D_1$ be the average distance between ancestors and the other viruses at time $t$, and $D_2$, the distance between the exposed and the other viruses. Note that an ancestor may recover and, therefore, not mutate in this time step. The Offspring term is a sum of different contributions between offspring and the other viruses in the population, as explained in detail below.

1. *Genetic distance between offspring and recovered.* The number of pairs is $x_t R_t$. Because offspring do not evolve in the time step they appear, their average distance is $D_1$. Then, its contribution to the Offspring term is $\boldsymbol{x_t R_t D_1}$.

2. *Genetic distance between offspring and exposed.* The number of pairs is $x_t E_t$. Because the exposed evolve, these pairs contribute with $\boldsymbol{x_t E_t (D_2 + B\mu)}$ to the Offspring term.

3. *Genetic Distance between offspring of an infected (ancestor) that does not recover (there are $(I_t - r_t)$ of these individuals) and infected*:

   (a) The distance between an offspring and its ancestor is $B\mu$, since the ancestor evolves. There are $x_t(I_t - r_t)/I_t$ new infections of this type, contributing with $\boldsymbol{x_t((I_t - r_t)/I_t)B\mu}$ to the distance.

   (b) For each offspring there are $I_t - r_t - 1$ infected individuals that did not recover and are not its ancestral. The distance between the offspring and these individuals is $(D_1 + B\mu)$, adding $\boldsymbol{x_t((I_t - r_t)/I_t)(I_t - r_t - 1)(D_1 + B\mu)}$ to the Offspring term.

   (c) The distance between the offspring and individuals that recover is $D_1$, because neither of these viruses evolve in this time step. There are $x_t((I_t - r_t)/I_t)r_t$ pairs of these viruses, adding $\boldsymbol{x_t((I_t - r_t)/I_t)r_t D_1}$ to the Offspring term.

4. *Genetic distance between offspring of infected (ancestor) that recover in this iteration (there are $r_t$ of these individuals) and infected*:

(a) The distance between offspring and its ancestor is zero, because none of them evolve.

(b) The distance between the offspring and the other viruses of type is $D_1$. There are $x_t r_t / I_t$ new infections of this type, contributing $(x_t r_t / I_t)(r_t - 1) D_1$ to the Offspring term.

(c) The distance between offspring and the other infected individuals is $(x_t r_t / I_t)(I_t - r_t)(D_1 + B\mu)$, since the other infected viruses evolve..

5. *Genetic distance between offspring.* Because each ancestor gives rise to only one new infection, this distance equals $D_1$, and once there are $x_t(x_t - 1)/2$ pairs of offspring, this contribution is $(x_t(x_t - 1)/2) D_1$.

6. By summing everything up, we get

$$\begin{aligned}\text{Offspring} = & \ x_t R_t D_1 + x_t E_t (D_2 + B\mu) \\ & + x \frac{(I_t - r_t)}{I_t} B\mu + x_t \frac{(I_t - r_t)}{I_t}(I_t - r_t - 1)(D_1 + B\mu) + x_t \frac{(I_t - r_t)}{I_t} r_t D_1 \\ & + x_t \frac{r_t}{I_t} 0 + x_t \frac{r_t}{I_t}(r_t - 1) D_1 + x_t \frac{r_t}{I_t}(I_t - r_t)(D_1 + B\mu) \\ & + \frac{x_t(x_t - 1)}{2} D_1. \end{aligned} \quad (A5)$$

Putting all these terms together and defining $Z_t \equiv 2 Z'_t$ we obtain

$$\begin{aligned} d_{t+1} = \frac{1}{Z_t} & \left( d_t (R_t + E_t + I_t)(R_t + E_t + I_t - 1) \right. \\ & + x_t D_1 (x_t - 3 + 2R_t + 2I_t + 2E_t D_2 / D_1) \\ & \left. + 2B\mu (E_t + I_t - r_t)(E_t + I_t + R_t + x_t - 1) \right). \end{aligned} \quad (A6)$$

The reason for assigning the distance $D_1$ between infected and other viruses, instead of $d_t$, is that infected individuals represent only a fraction of the viruses in the population, and the distance between them and other viruses grows over time, therefore being above the average $d_t$. The same holds for the exposed individuals.

Although we were not able to analytically find an expression for $D_1$ and $D_2$, we can approximate them as follows. First we assume that $D_2 \approx D_1$. When the epidemic begins, all viruses are infected, so that $D_1 = d_t$. However, the ratio between infected and recovered individuals decreases to zero along the epidemic, making $D_1$ larger than $d_t$. Thus, to first order, it is possible to approximate $D_1 \approx d_t(1 + \epsilon)$, with $\epsilon$ a function of the number of recovered individuals, $R_t/(I_t + E_t + R_t)$ and the average number of mutations $B\mu$. Our simulations showed that the linear function $D_1 = d_t(1 + 2B\mu R_t/(I_t + E_t + R_t))$ works well (considering the parameters in Appendix A), leading to the theoretical result expressed by Eq.(2) from the main text.

## 2.3 Continuum Limit

To achieve the continuum limit we start by substituting $r_t = R_{t+1} - R_t$ and $x_t = E_{t+1} - E_t + I_{t+1} - I_t + R_{t+1} - R_t$ in Eq.(2) from the main text and subtracting $d_t$ from both sides of this equation:

$$\begin{aligned} d_{t+1} - d_t = \frac{1}{Z_t} & \left\{ 2 d_t \left( E_{t+1} - E_t + I_{t+1} - I_t + R_{t+1} - R_t \right) \times \right. \\ & \times \left[ -1 + B\mu \frac{R_t}{I_t + E_t + R_t}(R_{t+1} + R_t + I_{t+1} + I_t + E_{t+1} + E_t - 3) \right] \\ & \left. + 2B\mu \left( E_t + I_t + R_t - R_{t+1} \right) \left( E_{t+1} + I_{t+1} + R_{t+1} - 1 \right) \right\} \end{aligned} \quad (A7)$$

with

$$Z_t = (E_{t+1} + I_{t+1} + R_{t+1})(E_{t+1} + I_{t+1} + R_{t+1} - 1). \tag{A8}$$

Then, we consider the first order approximations

$$f_t \approx f(t)$$
$$f_{t+1} \approx f(t) + \dot{f}(t)\Delta t,$$

and once $B\mu$ in the last line of Eq.(A7) is the number of mutations per time step, we replace it by $B\mu\Delta t$

$$\dot{d}(t)\Delta t = \tag{A9}$$
$$\frac{1}{Z_t}\left\{2d(t)\Delta t\left(\dot{E}(t) + \dot{I}(t) + \dot{R}(t)\right) \times \right.$$
$$\times\left[-1 + B\mu\frac{R(t)}{I(t) + E(t) + R(t)}\left(2R(t) + 2I(t) + 2E(t) + \Delta t(\dot{E}(t) + \dot{I}(t) + \dot{R}(t)) - 3\right)\right]$$
$$\left. + 2B\Delta t\mu\left(E(t) + I(t) - \dot{R}(t)\Delta(t)\right)\left(R(t) + I(t) + E(t) + \Delta t(\dot{E}(t) + \dot{I}(t) + \dot{R}(t)) - 1\right)\right\}$$
$$\tag{A10}$$

with

$$Z_t = (R(t) + I(t) + E(t) + \Delta t(\dot{E}(t) + \dot{I}(t) + \dot{R}(t))) \times$$
$$\times (R(t) + I(t) + E(t) + \Delta t(\dot{E}(t) + \dot{I}(t) + \dot{R}(t)) - 1). \tag{A11}$$

Finally, by taking the limit $\Delta t \to 0$ we obtain the continuous time equation.

## 2.4 Multiple Infections

The average distance $d_{root,t}^{(i)}$ between viruses from a lineage and its root is calculated using the same technique discussed above, however it is much simpler, once we only need to calculate the average distance from a kind of virus and the root (a single virus which does not evolve). Using the same notation, but now with a super-index to denote the lineage, we obtain

$$d_{root,t+1}^{(i)} = \frac{1}{Z_t}\left[\left(R_t^{(i)} + E_t^{(i)} + I_t^{(i)}\right)d_{root,t}^{(i)} + E_t^{(i)}B\mu + \left(I_t^{(i)} - r_t^{(i)}\right)B\mu + x_t^{(i)}D_{1,root}^{(i)}\right] \tag{A12}$$

with $Z_t = (E_t^{(i)} + I_t^{(i)} + R_t^{(i)} + x_t^{(i)})$ and $D_{1,root}^{(i)}$ being the average distance between infected and the root, which is given (similarly to $D_1$) by

$$D_{1,root}^{(i)} = d_{root,t}^{(i)}\left(1 + 4B\mu\frac{R_t^{(i)}}{E_t^{(i)} + I_t^{(i)} + R_t^{(i)}}\right).$$

The factor 4 is a fit from numerical investigations. The continuum limit is obtained by subtracting $d_{root,t}^{(i)}$ from both sides of Eq.(6) from the main text, applying the continuous approximation for each epidemic curve and taking the limit $\Delta t \to 0$.

## 3 Appendix C: Real genetic evolution algorithm

In order to estimate the real (from genetic data) genetic evolution, we used 55 complete genome sequences collected in China [8]. First, these sequences were ordered and

numbered by its collection date and a matrix of genetic distances $d_{ij}$ between genomes $i$ and $j$ has been constructed. Each pair of sequences were alligned with the Needleman-Wunsch algorithm, with score $+1$ for match and $-1$ for mismatch [9]. Then, the distance between two genomes was computed counting the number of substitutions between the sequences, neglecting *indels*.

We defined a time window $\tau_W = 14 = \tau_0$ days. Thus, every genome collected within $\tau_W$ are considered infected, and the genomes collected before this time window are considered recovered. Now, we calculate the average distance among the infected $d_{I,t}$, recovered $d_{R,t}$ and among infected and recovered $d_{IR,t}$ at the time $t$. Fig.1 shows an example of a distance matrix with a specific time window. Finally, the average distance at time $t$ can be computed as

$$d_t = \frac{d_{I,t} I_t (I_t - 1) + 2 d_{IR,t} I_t R_t + d_{R,t} R_t (R_t - 1)}{(R_t + I_t)(R_t + I_t - 1)} \tag{A13}$$

where $I_t$ and $R_t$ are respectively given by $I(t)$ and $R(t)$ described evaluated in the supplemental material.

With this algorithm, we obtained 20 non-overlapping sets of infected genomes. One of these sets contained only one sequence and was not usable; a second set was too far from all other data and was also discarded. Thus, we were able to calculate 18 points (that appear in Fig.4 from the main text) with error bars given by the standard deviation of each set of distances (between infected, recovered and between infected and recovered) at each time $t$.

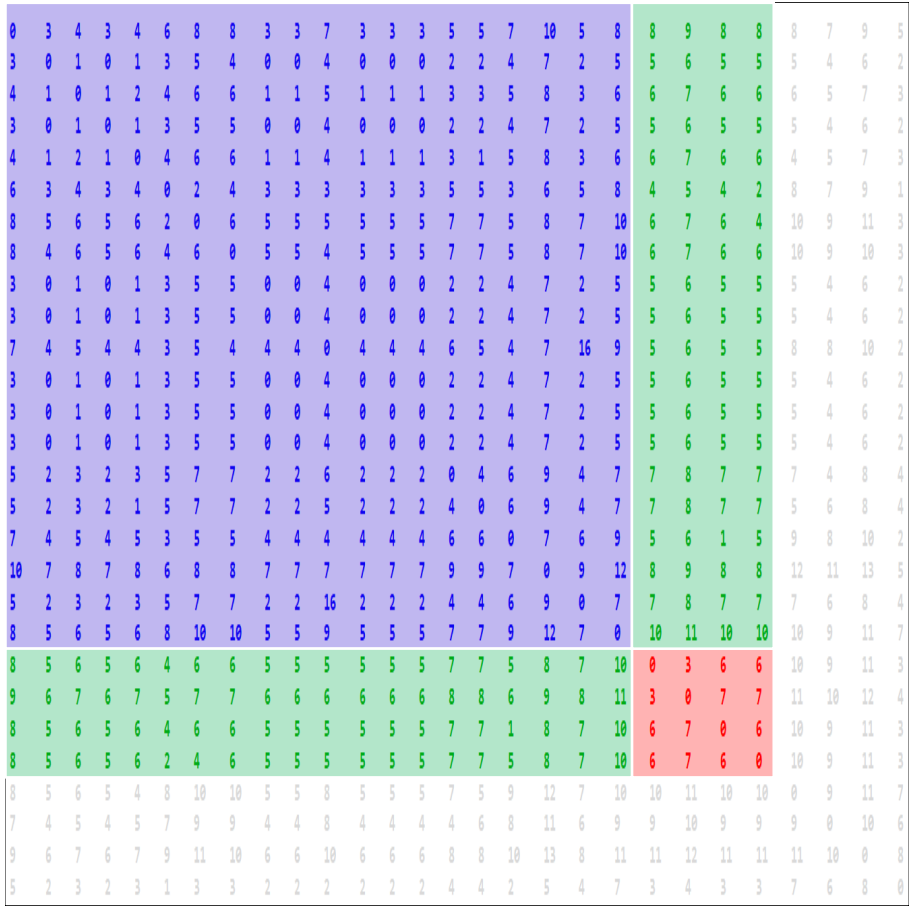

**Fig 1. Example of distance matrix to illustrate the algorithm to infer the genetic evolution.** Every genome collected within a time window $\tau_W$ is considered to belong to an infected individual. The red block shows distances between these viruses. The blue block shows viruses that appeared before the present time window, whose individuals are considered to have recovered. Green blocks are distances between infected and recovered individuals. The remaining entries are distances from viruses that have not appeared yet at that considered time, i.e., they appeared after the considered time window.

## 4 Appendix D: The COVID-19 data from China

We got the Chinese epidemic data from the dataset "Epidemic Data for Novel Coronavirus COVID-19" from Wolfram data repository [10]. Unfortunately, this dataset starts on 22 January (going up to 18 August by the date of our analysis), lacking the previous data. Another concern is about the change in the notification protocols adopted by the Chinese government. On 13 February, the Hubei province started to report not only the positive laboratory tests, but also the clinically diagnosed cases as infected too, appearing a sudden increase in infected curve [11]. We also need to correct the data by including undetected cases.

Firstly, in order to correct the notification problem, we smoothly distribute the sudden increase number of cases among the previous dates. Following reference [11], the

corrected accumulated number of cases $I_{a,c}(t)$ is given by

$$I_{a,c}(t) = I_a(t) + 15133 \frac{\sum_{i=22 \text{ Jan}}^{t} I_a(t)}{\sum_{i=22 \text{ Jan}}^{13 \text{ Feb}} I_a(t)} \quad (A14)$$

for $t \in \{22 \text{ Jan}, \ldots, 12 \text{ Feb}\}$, where $I_c(t)$ is the accumulated number of cases at date $t$, and $15133 = I_a(13 \text{ Feb}) - I_a(12 \text{ Feb})$ is the sudden increase due to the changes in the notification protocol.

Now, the undetected cases in China were estimated in reference [12], and also following reference [11], we get

$$I_{a,c'}(t) = \frac{I_{a,c}(t)}{1 - \theta(t)} \quad (A15)$$

for the estimated total number of cases at time $t$, where $\theta$ is the undetected fraction,

$$\theta(t) = \begin{cases} 0.86, \text{ for } t \leq 24 \text{ Jan} \\ linear\ decrease, \text{ for } 24 \text{ Jan} \leq t \leq 08 \text{ Feb} \\ 0.31, \text{ for } t \geq 08 \text{ Feb} \end{cases} \quad (A16)$$

This correction is also applied to the recovered curve. However, the Wolfram data distincts recovered $Rec(t)$ from deaths $Dea(t)$, while our theory does not differentiates these numbers. Thus, the number of recovered individuals we must consider is

$$R(t) = \frac{Rec(t) + Dead(t)}{1 - \theta(t)} \quad (A17)$$

and the infected curve is now given as

$$I(t) = I_{a,c'}(t) - R(t) \quad (A18)$$

Fig.2 shows the curves after these corrections. Once we do no have directly access to exposed data, we did not consider exposed individuals, meaning, at this point, that we are dealing with a SIR model without any prejudice to the present theory. However, bad data is an important source of error.

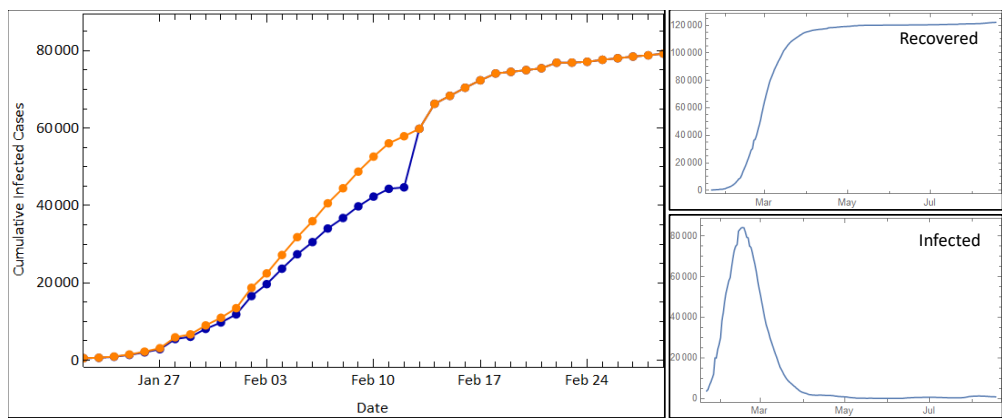

**Fig 2. Chinese epidemic curves after corrections.** The left chart shows the cumulative number of infections in China. The blue curve is the reported number of cases before the smoothness procedure of Eq.(14) and the orange curve is the result of this procedure. The right charts are the recovered and infected curves $R(t)$ and $I(t)$.

Finally, we fit an exponential curve to a few initial data points of $I(t)$ and $R(t)$ and extrapolate it to previous dates. For the $I$-curve, we have adjusted the exponential $e^{a(t-t_0)}$, with fit parameters $a$ and $t_0$, on the first $n_I = 10$ data points and extrapolated it up to the first case $t_0$ days before. With this approach, we found $t_0 = 11$ Dec, which is close to the first case reported by WHO, 08 Dec [13]. For the $R(t)$-curve, we have used the first $n_R = 13$ data points. The numbers $n_I$ and $n_R$ were chosen in order to make the exponential extrapolation makes sense according to WHO estimates of the first case, as also to make $R(t) < I(t)$ in a plausible way.

Now, the curves $R(t)$ and $I(t)$ can be implemented in the recurrence equation and the distance evolution can be estimated, with the first distance $d_0$ equalling zero.

**S1 Table. All Chinese genome sequences.** All genomes registered in Wolfram Repository "Genetic Sequences for the SARS-CoV-2 Coronavirus" with complete *NucleotideStatus* and human *Host* from China (data accessed 19/08/2020) [1].

| Accession Number | Collection Date | Length | Geographic Location | Included? | Justification |
|---|---|---|---|---|---|
| MN908947 | 26 Dec 2019* | 29903 | Wuhan, Hubei** | Y | |
| MN938384 | 10 Jan 2020 | 29838 | Shenzen, Guangdong | Y | |
| MN975262 | 11 Jan 2020 | 29891 | Wuhan, Hubei** | Y | |
| MN988668 | 02 Jan 2020 | 29881 | Wuhan, Hubei** | Y | |
| MN988669 | 02 Jan 2020 | 29881 | Wuhan, Hubei** | Y | |
| MN996527 | 30 Dec 2019 | 29825 | Wuhan, Hubei | Y | |
| MN996528 | 30 Dec 2019 | 29891 | Wuhan, Hubei | Y | |
| MN996529 | 30 Dec 2019 | 29852 | Wuhan, Hubei | Y | |
| MN996530 | 30 Dec 2019 | 29854 | Wuhan, Hubei | Y | |
| MN996531 | 30 Dec 2019 | 29857 | Wuhan, Hubei | Y | |
| MT019529 | 23 Dec 2019 | 29899 | Wuhan, Hubei | Y | |
| MT019530 | 30 Dec 2019 | 29889 | Wuhan, Hubei | N | MT19530 to MT19532: Might be biased data (sequences from the same researchers, collected at the same day and with quite the same length, with no other informations up to the date we have made the analysis).* |
| MT019531 | 30 Dec 2019 | 29899 | Wuhan, Hubei | N | |
| MT019532 | 30 Dec 2019 | 29890 | Wuhan, Hubei | N | |
| MT019533 | 01 Jan 2020 | 29883 | Wuhan, Hubei | Y | |
| MT034054 | 03 Jan 2020 | 29885 | Beijing | Y | |
| MT039873 | 20 Jan 2020 | 29833 | Hangzhou, Zhejiang | Y | |
| MT039874 | 22 Jan 2020 | 29858 | Hangzhou, Zhejiang** | Y | |
| MT049951 | 17 Jan 2020 | 29903 | Kunming,† Yunnan | Y | |
| MT079843 | 22 Jan 2020 | 29915 | Wuhan, Hubei** | Y | MT079843 to MT079854: Might be biased data (probable nosocomial transmission).** |
| MT079844 | 22 Jan 2020 | 29910 | Wuhan, Hubei** | N | |
| MT079845 | 22 Jan 2020 | 29955 | Wuhan, Hubei** | N | Then we have included only one genome. |
| MT079846 | 22 Jan 2020 | 29903 | Wuhan, Hubei** | N | |
| MT079847 | 22 Jan 2020 | 29872 | Wuhan, Hubei** | N | |
| MT079848 | 22 Jan 2020 | 29880 | Wuhan, Hubei** | N | |
| MT079849 | 22 Jan 2020 | 29904 | Wuhan, Hubei** | N | |
| MT079850 | 22 Jan 2020 | 29885 | Wuhan, Hubei** | N | |
| MT079851 | 22 Jan 2020 | 30018 | Wuhan, Hubei** | N | |
| MT079852 | 22 Jan 2020 | 29891 | Wuhan, Hubei** | N | |
| MT079853 | 22 Jan 2020 | 29766 | Wuhan, Hubei** | N | |
| MT079854 | 22 Jan 2020 | 29897 | Wuhan, Hubei** | N | |
| MT093631 | 08 Jan 2020 | 29860 | Beijing† | N | No detailed geographic information available. |
| MT121215 | 02 Feb 2020 | 29945 | Shanghai | Y | |

1

| Accession | Date | Length | Location | Included | Notes |
|---|---|---|---|---|---|
| MT123290 | 05 Feb 2020 | 29891 | Guangzhou, Guangdong | Y | |
| MT123291 | 29 Jan 2020 | 29882 | Guangzhou, Guangdong | Y | |
| MT123292 | 27 Jan 2020 | 29923 | Guangzhou, Guangdong | Y | |
| MT123293 | 29 Jan 2020 | 29871 | Guangzhou, Guangdong | Y | |
| MT135041 | 26 Jan 2020 | 29903 | Beijing | N | MT135041 to MT135044: Might be biased data (the lengths are all the same). |
| MT135042 | 28 Jan 2020 | 29903 | Beijing | N | |
| MT135043 | 28 Jan 2020 | 29903 | Beijing | N | Then we have included only one genome |
| MT135044 | 28 Jan 2020 | 29903 | Beijing | Y | |
| MT226610 | 20 Jan 2020 | 29899 | Kunming, Yunnan† | N | No detailed geographic information available. |
| MT253696 | 23 Jan 2020 | 29781 | Hangzhou, Zhejiang | N | MT253696 to MT253710: Might be biased data (cluster of cases*; also |
| MT253697 | 23 Jan 2020 | 29781 | Hangzhou, Zhejiang | N | they all have the same length). |
| MT253698 | 23 Jan 2020 | 29781 | Hangzhou, Zhejiang | N | Then we have included only one genome. |
| MT253699 | 24 Jan 2020 | 29781 | Hangzhou, Zhejiang | N | |
| MT253700 | 25 Jan 2020 | 29781 | Hangzhou, Zhejiang | N | |
| MT253701 | 21 Jan 2020 | 29781 | Hangzhou, Zhejiang | N | |
| MT253702 | 21 Jan 2020 | 29781 | Hangzhou, Zhejiang | N | |
| MT253703 | 25 Jan 2020 | 29781 | Hangzhou, Zhejiang | N | |
| MT253704 | 25 Jan 2020 | 29781 | Hangzhou, Zhejiang | N | |
| MT253705 | 22 Jan 2020 | 29781 | Hangzhou, Zhejiang | N | |
| MT253706 | 22 Jan 2020 | 29781 | Hangzhou, Zhejiang | N | |
| MT253707 | 25 Jan 2020 | 29781 | Hangzhou, Zhejiang | N | |
| MT253708 | 21 Jan 2020 | 29781 | Hangzhou, Zhejiang | N | |
| MT253709 | 21 Jan 2020 | 29781 | Hangzhou, Zhejiang | N | |
| MT253710 | 21 Jan 2020 | 29781 | Hangzhou, Zhejiang | Y | |
| MT259226 | 10 Jan 2020 | 29868 | Wuhan, Hubei | Y | |
| MT259227 | 26 Jan 2020 | 29863 | Wuhan, Hubei | Y | |
| MT259228 | 26 Jan 2020 | 29861 | Wuhan, Hubei | Y | |
| MT259229 | 26 Jan 2020 | 29864 | Wuhan, Hubei | Y | |
| MT259230 | 25 Jan 2020 | 29866 | Wuhan, Hubei | Y | |
| MT259231 | 25 Jan 2020 | 29865 | Wuhan, Hubei | Y | |
| MT281577 | 10 Mar 2020 | 29903 | Fujyang, Anhui | Y | |
| MT291826 | 30 Dec 2019 | 29807 | Wuhan, Hubei | Y | |
| MT291827 | 30 Dec 2019 | 29858 | Wuhan, Hubei | Y | |
| MT291828 | 30 Dec 2019 | 29858 | Wuhan, Hubei | Y | |
| MT291829 | 30 Dec 2019 | 29774 | Wuhan, Hubei | Y | |
| MT291830 | 30 Dec 2019 | 29807 | Wuhan, Hubei | Y | |
| MT291831 | 24 Jan 2020 | 29872 | Beijing | Y | |
| MT291832 | 25 Jan 2020 | 29828 | Beijing | Y | |

| | | | | | |
|---|---|---|---|---|---|
| MT291833 | 28 Jan 2020 | 29821 | | Beijing | Y |
| MT291834 | 28 Jan 2020 | 29865 | | Beijing | Y |
| MT291835 | 27 Jan 2020 | 29834 | | Beijing | Y |
| MT291836 | 29 Jan 2020 | 29860 | | Beijing | Y |
| MT407649 | 22 Jan 2020 | 29833 | | Hangzhou,[†] Zhejiang | Y |
| MT407650 | 22 Jan 2020 | 29821 | | Hangzhou,[†] Zhejiang | Y |
| MT407651 | 22 Jan 2020 | 29822 | | Hangzhou,[†] Zhejiang | Y |
| MT407652 | 26 Jan 2020 | 29835 | | Hangzhou,[†] Zhejiang | Y |
| MT407653 | 26 Jan 2020 | 29835 | | Hangzhou,[†] Zhejiang | Y |
| MT407654 | 24 Mar 2020 | 29817 | | Hangzhou,[†] Zhejiang | Y |
| MT407655 | 24 Mar 2020 | 29817 | | Hangzhou,[†] Zhejiang | Y |
| MT407656 | 24 Mar 2020 | 29835 | | Hangzhou,[†] Zhejiang | Y |
| MT407657 | 24 Mar 2020 | 29776 | | Hangzhou,[†] Zhejiang | Y |
| MT407658 | 24 Mar 2020 | 29770 | | Hangzhou,[†] Zhejiang | Y |
| MT407659 | 24 Mar 2020 | 29828 | | Hangzhou,[†] Zhejiang | Y |
| MT412134 | 24 Feb 2020 | 29867 | | Zhengzhou, Henan[†] | N | No detailed geographic information available. |
| MT446312 | 05 Feb 2020 | 29879 | | Guangzhou, Guangdong[†] | Y |
| MT510727 | 15 Feb 2020 | 29903 | | Meizhou, Guangdong[†] | N | MT510727 and MT510728: Might be biased data (data from familial |
| MT510728 | 13 Feb 2020 | 29903 | | Meizhou, Guangdong[†] | N | cluster*). There is also no detailed geographic information available. |
| MT534630 | 26 Jan 2020 | 29845 | | Changzhou,Jiangsu | Y |
| MT568634 | 25 Feb 2020 | 29861 | | Guangzhou, Guangdong | N | MT568634 to MT568641: data from a work presenting different approaches |
| MT568635 | 25 Feb 2020 | 29854 | | Guangzhou, Guangdong | N | for genome sequencing.** |
| MT568636 | 27 Feb 2020 | 29858 | | Guangzhou, Guangdong | N | Then, this data might have more errors than the others. |
| MT568637 | 25 Feb 2020 | 29860 | | Guangzhou, Guangdong | N |
| MT568638 | 25 Feb 2020 | 29854 | | Guangzhou, Guangdong | N |
| MT568639 | 25 Feb 2020 | 29861 | | Guangzhou, Guangdong | N |
| MT568640 | 25 Feb 2020 | 29858 | | Guangzhou, Guangdong | N |
| MT568641 | 25 Feb 2020 | 29868 | | Guangzhou, Guangdong | N |
| MT622319 | 23 Jan 2020 | 29889 | | Shanghai[†] | N | No detailed geographic information available. |
| MT627325 | 28 Feb 2020 | 29859 | | Shanghai[†] | N | No detailed geographic information available. |
| NC045512 | Dec 2019 | 29903 | | Wuhan, Hubei** | N | Identical to MN908947.* |

**S2 Table. Included sequences sorted by Collection Date.** All informations according to S1 Table.

| Number | Accession Number | Collection Date | Length | Geographic Location |
|---|---|---|---|---|
| #1 | MT019529 | 23 Dec 2019 | 29899 | Wuhan, Hubei |
| #2 | MN908947 | 26 Dec 2019 | 29903 | Wuhan, Hubei |
| #3 | MT291829 | 30 Dec 2019 | 29774 | Wuhan, Hubei |
| #4 | MT291826 | 30 Dec 2019 | 29807 | Wuhan, Hubei |
| #5 | MT291830 | 30 Dec 2019 | 29807 | Wuhan, Hubei |
| #6 | MN996527 | 30 Dec 2019 | 29825 | Wuhan, Hubei |
| #7 | MN996529 | 30 Dec 2019 | 29852 | Wuhan, Hubei |
| #8 | MN996530 | 30 Dec 2019 | 29854 | Wuhan, Hubei |
| #9 | MN996531 | 30 Dec 2019 | 29857 | Wuhan, Hubei |
| #10 | MT291827 | 30 Dec 2019 | 29858 | Wuhan, Hubei |
| #11 | MT291828 | 30 Dec 2019 | 29858 | Wuhan, Hubei |
| #12 | MN996528 | 30 Dec 2019 | 29891 | Wuhan, Hubei |
| #13 | MT019533 | 01 Jan 2020 | 29883 | Wuhan, Hubei |
| #14 | MN988668 | 02 Jan 2020 | 29881 | Wuhan, Hubei |
| #15 | MN988669 | 02 Jan 2020 | 29881 | Wuhan, Hubei |
| #16 | MT034054 | 03 Jan 2020 | 29885 | Beijing |
| #17 | MN938384 | 10 Jan 2020 | 29838 | Shenzhen, Guangdong |
| #18 | MT259226 | 10 Jan 2020 | 29868 | Wuhan, Hubei |
| #19 | MN975262 | 11 Jan 2020 | 29891 | Wuhan, Hubei |
| #20 | MT049951 | 17 Jan 2020 | 29903 | Yunnan |
| #21 | MT039873 | 20 Jan 2020 | 29833 | Hangzhou, Zhejiang |
| #22 | MT253710 | 21 Jan 2020 | 29781 | Hangzhou, Zhejiang |
| #23 | MT407650 | 22 Jan 2020 | 29821 | Zhejiang |
| #24 | MT407651 | 22 Jan 2020 | 29822 | Zhejiang |
| #25 | MT407649 | 22 Jan 2020 | 29833 | Zhejiang |
| #26 | MT039874 | 22 Jan 2020 | 29858 | Hangzhou, Zhejiang |
| #27 | MT079843 | 22 Jan 2020 | 29915 | Wuhan, Hubei |
| #28 | MT291831 | 24 Jan 2020 | 29872 | Beijing |
| #29 | MT291832 | 25 Jan 2020 | 29828 | Beijing |
| #30 | MT259231 | 25 Jan 2020 | 29865 | Wuhan, Hubei |
| #31 | MT259230 | 25 Jan 2020 | 29866 | Wuhan, Hubei |
| #32 | MT407652 | 26 Jan 2020 | 29835 | Zhejiang |
| #33 | MT407653 | 26 Jan 2020 | 29835 | Zhejiang |
| #34 | MT534630 | 26 Jan 2020 | 29845 | Changzhou, Jiangsu |
| #35 | MT259228 | 26 Jan 2020 | 29861 | Wuhan, Hubei |
| #36 | MT259227 | 26 Jan 2020 | 29863 | Wuhan, Hubei |
| #37 | MT259229 | 26 Jan 2020 | 29864 | Wuhan, Hubei |
| #38 | MT291835 | 27 Jan 2020 | 29834 | Beijing |
| #39 | MT123292 | 27 Jan 2020 | 29923 | Guangzhou, Guangdong |
| #40 | MT291833 | 28 Jan 2020 | 29821 | Beijing |
| #41 | MT291834 | 28 Jan 2020 | 29865 | Beijing |
| #42 | MT135044 | 28 Jan 2020 | 29903 | Beijing |
| #43 | MT291836 | 29 Jan 2020 | 29860 | Beijing |
| #44 | MT123293 | 29 Jan 2020 | 29871 | Guangzhou, Guangdong |
| #45 | MT123291 | 29 Jan 2020 | 29882 | Guangzhou, Guangdong |
| #46 | MT121215 | 02 Feb 2020 | 29945 | Shanghai |
| #47 | MT446312 | 05 Feb 2020 | 29879 | Guangzhou, Guangdong |
| #48 | MT123290 | 05 Feb 2020 | 29891 | Guangzhou, Guangdong |
| #49 | MT281577 | 10 Mar 2020 | 29903 | Fuyang, Anhui |
| #50 | MT407658 | 24 Mar 2020 | 29770 | Zhejiang |
| #51 | MT407657 | 24 Mar 2020 | 29776 | Zhejiang |
| #52 | MT407654 | 24 Mar 2020 | 29817 | Zhejiang |
| #53 | MT407655 | 24 Mar 2020 | 29817 | Zhejiang |
| #54 | MT407659 | 24 Mar 2020 | 29828 | Zhejiang |
| #55 | MT407656 | 24 Mar 2020 | 29835 | Zhejiang |

**S3 Table. Genome information used to calculate points in Fig.5.** We have used a 14 days time window, i. e., every sequenced genome within an interval of 14 days were considered as infected ones, while the previous were considered to be recovered.

| Point Number* | Infected Genomes | Recovered Genomes | Date Interval |
|---|---|---|---|
| #1 | #03 → #19 | #1 → #02 | 30 Dec 2019 → 12 Jan 2020 |
| #2 | #13 → #19 | #1 → #12 | 01 Jan 2020 → 14 Jan 2020 |
| #3 | #14 → #19 | #1 → #13 | 02 Jan 2019 → 15 Jan 2020 |
| #4 | #16 → #19 | #1 → #15 | 03 Jan 2019 → 16 Jan 2020 |
| #5 | #17 → #27 | #1 → #16 | 10 Jan 2019 → 23 Jan 2020 |
| #6 | #19 → #28 | #1 → #18 | 11 Jan 2019 → 24 Jan 2020 |
| #7 | #20 → #45 | #1 → #19 | 17 Jan 2019 → 30 Jan 2020 |
| #8 | #21 → #46 | #1 → #20 | 20 Jan 2019 → 02 Feb 2020 |
| #9 | #22 → #46 | #1 → #21 | 21 Jan 2019 → 03 Feb 2020 |
| #10 | #23 → #46 | #1 → #22 | 22 Jan 2019 → 04 Feb 2020 |
| #11 | #28 → #48 | #1 → #27 | 24 Jan 2019 → 06 Feb 2020 |
| #12 | #29 → #48 | #1 → #28 | 25 Jan 2019 → 07 Feb 2020 |
| #13 | #32 → #48 | #1 → #31 | 26 Jan 2019 → 08 Feb 2020 |
| #14 | #38 → #48 | #1 → #37 | 27 Jan 2019 → 09 Feb 2020 |
| #15 | #40 → #48 | #1 → #39 | 28 Jan 2019 → 10 Feb 2020 |
| #16 | #43 → #48 | #1 → #42 | 29 Jan 2019 → 11 Feb 2020 |
| #17 | #46 → #48 | #1 → #45 | 02 Feb 2019 → 15 Feb 2020 |
| #18 | #47 → #48 | #1 → #46 | 05 Feb 2019 → 18 Feb 2020 |
| #19** | #49 → #49 | #1 → #48 | |
| #20† | #50 → #55 | #1 → #49 | 24 Mar 2019 → 06 Apr 2020 |

*In Fig.4 from the main text, points are numbered from left to right.

**Since there is only one genome in this time window, we cannot estimate a distance among the infected population, so genome #49 was not used.

† This point was not included in Fig.5 because it is lacking more than one month of genetic information between points #18 and #19, therefore the distance among the recovered population cannot be well inferred.



\end{document}